\documentclass[12pt]{article}
\usepackage[T2A]{fontenc}
\usepackage{latexsym,amsthm,amssymb,amsfonts,amsmath,amscd,wrapfig}
\usepackage{graphicx}
\usepackage{citehack}
\usepackage{showkeys}

\makeatletter \@addtoreset{equation}{section} \makeatother

\newtheorem{proposition}{Proposition}
\newtheorem{theorem}{Theorem}
\newtheorem{lemma}{Lemma}

\def\dfrac{\displaystyle\frac}
\def\sumd{\displaystyle\sum}

\def\intd{\displaystyle\int}

\begin{document}
\title{On Universality of Bulk Local Regime of the Deformed Laguerre
Ensemble}
\author{ T. Shcherbyna\\
 Institute for Low Temperature Physics, Kharkov,
Ukraine. \\E-mail: t\underline{ }shcherbina@rambler.ru
}
\date{}

\maketitle \centerline{{\it 2000 Mathematics Subject Classification.} Primary 15A52; Secondary 15A57}

\maketitle
\begin{abstract}
We consider the deformed Laguerre Ensemble $H_n=\dfrac{1}{m}\Sigma_n^{1/2}A_{m,n}A_{m,n}^*\Sigma_n^{1/2}$ in which
$\Sigma_n$ is a
positive hermitian matrix (possibly random) and $A_{m,n}$ is a $n\times m$ complex Gaussian random matrix (independent of
$\Sigma_n$), $\dfrac{m}{n}\to c>1$. Assuming that the Normalized Counting Measure of $\Sigma_n$ converges
weakly (in probability) to a non-random measure $N^{(0)}$ with a bounded support we prove the universality of
the local eigenvalue statistics in the bulk of the limiting spectrum of $H_n$.
\end{abstract}
\section{Introduction.}
Universality is an important topic of the random matrix theory. It
deals with statistical properties of eigenvalues of $n\times n$
random matrices on intervals whose length tends to zero as $n \to
\infty$. According to the universality conjecture these properties
do not depend to large extent on the ensemble. The conjecture was
proposed by Dyson in the early 60s.
To formulate it we need some notations
and definitions. Denote by
$\lambda_1^{(n)},\ldots,\lambda_n^{(n)}$ the eigenvalues of the
random matrix. Define the normalized eigenvalue counting measure
(NCM) of the matrix as
\begin{equation}  \label{NCM}
N_n(\triangle)=\sharp\{\lambda_j^{(n)}\in
\triangle,j=\overline{1,n} \}/n,\quad N_n(\mathbb{R})=1,
\end{equation}
where $\triangle$ is an arbitrary interval of the real axis. For many known
random matrices the expectation $\overline{N}_n=\mathbf{E}\{N_n\}$ is
absolutely continuous, i.e.,
\begin{equation}  \label{rhon}
\overline{N}_n(\triangle)=\displaystyle\int\limits_\triangle \rho_n
(\lambda)d\,\lambda.
\end{equation}
The non-negative function $\rho_n$ in (\ref{rhon}) is called the density of
states.

Define also the $k$-point correlation function $R_k^{(n)}$ by the
equality:
\begin{equation}  \label{R}
\mathbf{E}\left\{ \sum_{j_{1}\neq ...\neq j_{k}}\varphi_k
(\lambda_{j_{1}},\dots,\lambda_{j_{k}})\right\} =\int \varphi_{k}
(\lambda_{1},\dots,\lambda_{k})R_{k}^{(n)}(\lambda_{1},\dots,\lambda_{k})
d\lambda_{1},\dots,d\lambda_{k},
\end{equation}
where $\varphi_{k}: \mathbb{R}^{k}\rightarrow \mathbb{C}$ is
bounded, continuous and symmetric in its arguments and the
summation is over all $k$-tuples of distinct integers $
j_{1},\dots,j_{k}\in\{1,\ldots,n\}$. Here and below integrals
without limits denote the integration over the whole real axis.

The behavior of $N_n$ as $n\to\infty$ is studied for many ensembles. It is shown that
$N_n$ converges weakly to a non-random limiting measure $N$. The limiting measure is normalized
to unity and as a rule is absolutely continuous
\begin{equation}  \label{rho}
N(\mathbb{R})=1,\quad N(\triangle)=\displaystyle\int\limits_\triangle
\rho(\lambda)d\,\lambda.
\end{equation}
The non-negative function $\rho$ in (\ref{rho}) is called the limiting
density of states of the ensemble.

We will call the spectrum the support of $N$ and define the bulk
of the spectrum as
\begin{equation*}
\hbox{bulk}\,N=\{\lambda|\exists (a,b)\subset \hbox{supp}\,N:\lambda\in (a,b),\,\,
\inf\limits_{\mu\in (a,b)}\rho(\mu)>0\}.
\end{equation*}
Then the universality hypothesis on the bulk of the spectrum says
that for $ \lambda_0\in\hbox{bulk}\, N$ we have:

(i) for any fixed $k$ uniformly in $\xi_1, \xi_2,\ldots, \xi_k$ varying
in any compact set in $\mathbb{R}$
\begin{equation}  \label{Un}
\lim\limits_{n\to \infty}\displaystyle\frac{1}{(n\rho_n(\lambda_0))^k}
R^{(n)}_k\left(\lambda_0+\displaystyle\frac{\xi_1}{\rho_n(\lambda_0)\,n},
\ldots,\lambda_0+\displaystyle\frac{\xi_k}{\rho_n(\lambda_0)\,n}\right)=\det \{S(\xi_i-\xi_j)\}_{i,j=1}^k,
\end{equation}
where
\begin{equation}  \label{S}
S(x)=\displaystyle\frac{\sin \pi x}{\pi x},
\end{equation}
and $R^{(n)}_k$ and $\rho_n$ are defined in (\ref{R}) and (\ref{rhon});

(ii) if
\begin{equation}
E_{n}(\triangle )=\mathbf{P}\{\lambda_{i}^{(n)}\not\in \triangle
,\,i= \overline{1,n}\},  \label{gapp}
\end{equation}
is the gap probability, then
\begin{equation}
\lim\limits_{n\rightarrow \infty }E_{n}\left( \left[\lambda_{0}+
\displaystyle \frac{a}{\rho_{n}(\lambda _{0})\,n},\lambda_{0}
+\displaystyle\frac{b}{\rho_{n}(\lambda_{0})\,n}\right]\right) =\det \{1-S_{a,b}\},  \label{gp}
\end{equation}
where the operator $S_{a,b}$ is defined on $L_{2}[a,b]$ by the formula
\begin{equation*}
S_{a,b}f(x)=\displaystyle\int\limits_{a}^{b}S(x-y)f(y)d\,y,
\end{equation*}
and $S$ is defined in (\ref{S}).

Bulk universality was proved initially for ensembles with Gaussian entries (see \cite{Me:91}). Then at the
end of 90's it was proved for unitary invariant ensembles of random matrices
(known also as unitary matrix models)(\cite{Pa-Sh:97,De-Co:99, Pa-Sh:07}), end then
for Wigner ensemble with some special distribution (see \cite{Jo:01}).
Recently in the series of important papers \cite{TaoVu:09}-\cite{ErTao:09} it was shown that local eigenvalue
statistics for matrices with independent entries depends only on the first few moments of the distribution of
entries. Hence, to prove the universality conjecture it suffices to prove it for the corresponding models
with the same moments of the distribution of entries (e.g. for gaussian entries). These results solved the
universality problem for Wigner ensemble with subexponential decay of entries.

In this paper we consider
the deformed Laguerre Ensemble, i.e. $n\times n$ matrices
\begin{equation}\label{H}
H_n=\dfrac{1}{m}\Sigma_n^{1/2}A_{m,n}A_{m,n}^*\Sigma_n^{1/2},
\end{equation}
where $\Sigma_{n}$ is a positive Hermitian $n\times n$ matrix with eigenvalues
$\{t_{j}^{(n)}\}_{j=1}^{n}\subset \mathbb{R}$ and $A_{m,n}$ is a $n\times m$ matrix, whose entries  $\Re
a_{\alpha j}$ and $\Im a_{\alpha j}$ are independent Gaussian random variables such that
\begin{equation}\label{A}
{\bf E}\{\Re a_{\alpha j}\}={\bf E}\{\Im a_{\alpha j}\}=0,\quad {\bf E}\{\Re^2 a_{\alpha j}\}={\bf E}\{\Im^2
a_{\alpha j}\}= \dfrac{1}{2},\quad \alpha=\overline{1,n},\,j=\overline{1,m},
\end{equation}
moreover
\begin{equation}\label{c}
c_{m,n}:=\dfrac{m}{n}\to c,\quad m,n\to\infty.
\end{equation}
Let
\begin{equation*}
N_{n}^{(0)}(\triangle )=\sharp \{t_{j}^{(n)}\in \triangle ,j=\overline{1,n} \}/n.
\end{equation*}
be the Normalized Counting Measure of eigenvalues of $\Sigma_{n}$.


The behavior of NCM for the ensemble (\ref{H}) -- (\ref{A}) is studied well enough. Indeed, it is easy
to see that the spectral properties of matrices (\ref{H}) and of the $m\times m$ matrices $m^{-1}A_{m,n}^*\Sigma_nA_{m,n}$
are closely related. For example, if $N_n$ and $N_m^*$ are the Normalized Counting Measures of (\ref{NCM}) of
these ensembles, then we have
\begin{equation}\label{sv}
N_m^*=(1-c_{m,n}^{-1})\delta_0+c_{m,n}^{-1}N_n,
\end{equation}
where $\delta_0$ is the Dirac $\delta$-function.
For matrices $m^{-1}A_{m,n}^*\Sigma_nA_{m,n}$ it was shown
in \cite{Mar-Pa:67} that if $N_{n}^{(0)}$ converges weakly with probability 1 to a non-random measure $N^{(0)}$ as
$n\rightarrow \infty $, then $N_{m}^*$ also converges weakly with probability 1 to a non-random measure $N^*$.
Moreover the Stieltjes transforms $f^*$ of $N^*$ satisfies the equation
\[
f^*(z)=-\left(z-c^{-1}\displaystyle\int\dfrac{tN^{(0)}(d\,t)}{1+tf(z)}\right)^{-1}.
\]
Hence, we have that if $N_{n}^{(0)}$ converges weakly
to a non-random measure $N^{(0)}$ as $n\rightarrow \infty $, then $N_{n}$ for matrices (\ref{H}) also converges weakly
to a non-random measure $N$, and since according to (\ref{sv}) we have that
\[
f^*(z)=\dfrac{1-c^{-1}}{z}+c^{-1}f(z),
\]
where $f$ is the Stieltjes transforms of $N$, we obtain the equation
\begin{equation}\label{eqv_f}
c^{-1}f(z)+\dfrac{1-c^{-1}}{z}=-\left(z-c^{-1}\displaystyle\int\dfrac{tN^{(0)}(d\,t)}{1+t(c^{-1}f(z)+(1-c^{-1})/z)}\right)^{-1}.
\end{equation}
The result on the local regime for the models of the type (\ref{H}) are much more pure. In the case $\Sigma_n=1$
the universality of the local regime (in the bulk and near the edges of the spectrum) was studied in \cite{NW:92}.
The bulk universality for the case $\Sigma_n=1$, $m/n\to 1$, but $A_{m,n}=A_{m,n}^{(0)}+A_{m,n}^{(1)}$,
with gaussian $A_{m,n}^{(0)}$ and $A_{m,n}^{(1)}$ with i.i.d., but not necessary gaussian entries, was studied
in \cite{BenPe:05}. In \cite{BaBenPe:05} the limiting distribution of the largest eigenvalue for ensemble
(\ref{H}) with $\Sigma_n=I+P$, where $P$ is a finite rank operator, was investigated.

In the present paper we prove universality of the local bulk regime for random matrices (\ref{H}) for a rather
general class of $\Sigma_n$. The main result is the following theorem.


\begin{theorem}
\label{thm:1} Let $c<1$ and the eigenvalues $\{t_{j}^{(n)}\}_{j=1}^{n}\subset (\mathbb{R}^+)^n$ of \,
$\Sigma_{n}$ in (\ref{H}) be a collection of random variables independent of $A_{m,n}$ of (\ref{A}).
Assume that there
exists a non-random measure $N^{(0)}$ of a bounded support $\sigma\in \mathbb{R}^+$ such
that  for any finite interval $ \Delta \subset \mathbb{R}$ and for
any $\varepsilon >0$
\begin{equation}\label{conpN0}
\lim_{n\rightarrow \infty }\mathbf{P}^{(\tau)}_n\{|N^{(0)}(\Delta )-N_{n}^{(0)}(\Delta )|>\varepsilon \}=0.
\end{equation}
Then for any $\lambda_{0}\in \emph{bulk}\,N$ the universality
properties (\ref{Un}) and (\ref{gp}) hold.
\end{theorem}

The paper is organized as follows. In Section $2$ we give a proof of determinant formulas for correlation functions
(\ref{R}) by the method from those of \cite{Br-Hi:96,Br-Hi:97}. This formula coincide with formula for the kernel
$K_{M,N}$ in \cite[Prop.2.1]{BaBenPe:05}, where the limiting distribution of the largest eigenvalue for ensemble
(\ref{H}) for $\Sigma_n=I+P$ with finite rank perturbation $P$ was investigated. Theorem \ref{thm:1} will be
proved in Section $3$. The method which is used for the limiting transition for the kernel is similar to
that of \cite{TSh:08}. Section $4$ deals with the proof of auxiliary statements for Theorem~\ref{thm:1}.


Note that we denote by $C, C_1$, etc. various constants appearing below, which
can be different in different formulas.

\section{The determinant formulas.}
It is well known (see, e.g., \cite{Me:91}) that the
correlation functions (\ref{R}) for the GUE $n\times n$ matrix can be written in the
determinant form
\begin{equation}
R_k^{(n)}(\lambda_1,\ldots,\lambda_k)=\det\{K_n(\lambda_i,\lambda_j)\}_{i,j=1}^k
\end{equation}
with
\begin{equation*}
K_n(\lambda_i,\lambda_j)=\sum\limits_{s=0}^{n-1}\phi_s(\lambda_i)\phi_s(
\lambda_j), \quad \phi_s(x)=n^{1/4}h_s(\sqrt{n}x)e^{-nx^2/4},
\end{equation*}
where $\{h_s\}_{s\ge 0}$ are orthonormal Hermite polynomials.
We want to find analogs of these formulas in the case of random
matrices (\ref{H}). To do this we will use the method from \cite{Br-Hi:96, Br-Hi:97},
where the determinant formulas for deformed Gaussian Unitary Ensemble were derived.
\begin{proposition}\label{p:ker}
Let $H_n$ be the random matrix defined in (\ref{H}) and $R_k^{(n)}$ be the correlation function (\ref{R}).
Then we have
\begin{equation}  \label{Det}
R_k^{(n)}(\lambda_1,\ldots,\lambda_k)=\mathbf{E}_n^{(\tau)}\{\det\{K_n(\lambda_i,\lambda_j)\}_{i,j=1}^k\},
\end{equation}
with
\begin{equation}\label{K}
K_n(\lambda,\mu)=-\dfrac{m}{4\pi^2}\oint\limits_L\oint\limits_\omega \dfrac{\exp\left\{ m(\lambda u-\mu
t)\right\}t^m}{(u-t)u^m} \prod\limits_{j=1}^n\left(\dfrac{u-\tau_j} {t-\tau_j}\right)d\,t d\,u,
\end{equation}
where $\tau_j=1/t_j^{(n)}$, $L$ is a closed contour, encircling $\{\tau_j\}_{j=1}^n$ and $\omega$ is any
closed contour encircling $0$ and not intersecting $L$.
\end{proposition}
The symbol $\mathbf{E}_n^{(\tau)}\{\ldots \}$ here and below denotes the expectation with respect to the
measure generated by $\Sigma_{n}$.
\begin{proof}
The probability distribution $P_n(H_n)$ for ensemble (\ref{H}) is given by (see,e.g.,\cite{BaBenPe:05})
\begin{equation}\label{p_h}
P_n(H_n)=\dfrac{1}{Z_n}e^{-m\hbox{tr}\,(\Sigma_n^{-1}H_n)}\hbox{det}^{m-n} H_n,
\end{equation}
where
\begin{equation}\label{Z}
Z_n=\int e^{-m\hbox{tr}\,(\Sigma_n^{-1}H_n)}\hbox{det}^{m-n} H_n d\, H_n.
\end{equation}

 Let us first calculate $Z_n$. Set
\begin{equation}\label{change}
\Sigma_n^{-1}=V_0^*LV_0,\quad H_n=V^*XV
\end{equation}
where $L=\hbox{diag}\,(\tau_1,\ldots,\tau_n)$, $\tau_j=1/t_j^{(n)}$, $X=\hbox{diag}\,(x_1,\ldots,x_n)$ and
$V_0$, $V$ are the matrices diagonalizing $\Sigma_n^{-1}$ and $H_n$ correspondingly. Then the differential
$d\, H_n$ in (\ref{Z}) transforms to $\triangle^2(X)d\,X d\,\mu(V)$, where $d\,X=\prod\limits_{j=1}^nd\,x_j$,
\begin{equation}\label{VdM}
\triangle(X)=\prod\limits_{i<j}^n (x_i-x_j)
\end{equation}
is a Vandermonde determinant, and $\mu(V)$ is the normalized to unity Haar measure on the unitary group
$U(n)$. Integral over the unitary group $U(n)$ can be easily computed using the well-known Harish
Chandra/Itsykson-Zuber formula (see \cite{Me:91}, Appendix 5)
\begin{proposition}\label{p:Its-Z}
Let $A$ and $B$ be normal $n\times n$ matrices with eigenvalues
$\{a_i\}_{i=1}^n$, $\{b_i\}_{i=1}^n$, correspondingly. Then we
have
\begin{equation}\label{Its-Zub}
\int \exp\{\hbox{tr} AU^*BU\} d\,\mu(U)=\dfrac{\det[\exp\{a_ib_j\}]_{i,j=1}^n}{\triangle(A)\triangle(B)},
\end{equation}
where $\triangle(A)$ and $\triangle(B)$ are Vandermonde determinants for the eigenvalues of $A$ and $B$.
\end{proposition}
Shifting $VV_0^*\to V$ and using (\ref{Its-Zub}) we have from (\ref{Z})
\begin{equation}
Z_n=\int\limits_0^\infty \dfrac{\det\{e^{-m\tau_jx_k}\}_{j,k=1}^n\triangle(X)}
{\triangle(L)}\prod\limits_{j=1}^nx_j^{m-n}d\,X.
\end{equation}
Since the function under the integral here is symmetric of $\{x_j\}_{j=1}^n$ and $m>n$, we obtain
\begin{equation}\label{Z_n}
\begin{array}{c}
Z_n=\dfrac{n!}{\triangle(L)}\int\limits_0^\infty
e^{-m\sum\limits_{j=1}^n\tau_jx_j}\triangle(X)\prod\limits_{j=1}^nx_j^{m-n}d\,X\\
=\dfrac{n!}{\triangle(L)}\det\left\{\int\limits_0^\infty
e^{-m\sum\limits_{j=1}^n\tau_jx_j}x_j^{m-n+l-1}d\,x_j\right\}_{j,l=1}^n
=\frac{n!m^{n(n-1)/2}\prod\limits_{l=1}^n(m-n+l-1)!}{\prod\limits_{j=1}^n(m\tau_j)^m}.
\end{array}
\end{equation}
Let us consider the function
\begin{equation*}
U_k(t_1,\ldots,t_k)={\bf E}\{\hbox{tr}\,e^{it_1H_n}\ldots \hbox{tr}\, e^{it_kH_n}\}.
\end{equation*}
Substituting expression (\ref{p_h}) we obtain
\begin{equation}\label{U}
U_k(t_1,\ldots,t_k)=\dfrac{1}{Z_n}\int e^{-m\hbox{tr}\,(\Sigma_n^{-1}H_n)}\hbox{det}^{m-n} H_n
\hbox{tr}\,e^{it_1H_n}\ldots \hbox{tr}\, e^{it_kH_n}\,d\,t_1\ldots d\,t_k.
\end{equation}
According to the definition of $R_k^{(n)}$ we have
\[
R_k^{(n)}(\lambda_1,\ldots,\lambda_k)=\dfrac{1}{Z_n}\int d\,t_1\ldots d\,t_k\, e^{-i\lambda_1t_1-\ldots
-i\lambda_kt_k}U_k(t_1,\ldots,t_k).
\]
Doing the change (\ref{change}) and using (\ref{Its-Zub}) we get
\begin{eqnarray}\label{R_1}\notag
R_k^{(n)}(\lambda_1,\dots,\lambda_k)&=&\dfrac{1}{Z_n\Delta(L)}\prod_{s=1}^k\left(\int
e^{-i\lambda_st_s}d\,t_s\right)
\prod_{j=1}^n\left(\int\limits_{0}^\infty x_j^{m-n}d\,x_j\right)\\
&&\times\det\left\{e^{-mx_j\tau_l}\right\}_{j,l=1}^n\Delta(X)
\prod\limits_{s=1}^k\left(\sum\limits_{j=1}^ne^{it_sx_j}\right).
\end{eqnarray}
Since the integral in (\ref{R_1}) is symmetric function of $\{x_l\}_{l=1}^n$, we can write
\begin{eqnarray*}
R_k^{(n)}(\lambda_1,\dots,\lambda_k)&=&\dfrac{n!}{Z_n\triangle(L)}\sum\limits_{\alpha_1,\dots,\alpha_k=1}^n
\prod_{s=1}^k\left(\int d\,t_s\right)\\
&&\prod_{j=1}^n\left(\int\limits_{0}^\infty x_j^{m-n}d\,x_j\right)
 e^{-m\sum\limits_{j=1}^n x_j\tau_j+\sum\limits_{s=1}^kit_s(x_{\alpha_s}-\lambda_s)}\Delta(X)
\end{eqnarray*}
It is easy to see, that if some of $\alpha_j$'s coincide (for example, $\alpha_1=\alpha_2=\ldots=\alpha_l$), then
the integral over $t_1,\dots,t_l$ in such term becomes $\delta(x_{\alpha_1}-\lambda_1)\prod\limits_{i=1}^{l-1}
\delta(\lambda_i-\lambda_{i+1})$
and hence can be omitted for $\lambda_i\not=\lambda_j$. Therefore, integrating over $t_s$
for $\lambda_i\ne\lambda_j$ we obtain
\begin{eqnarray*}
R_k^{(n)}(\lambda_1,\ldots,\lambda_k) &=&\dfrac{n!}{Z_n\triangle(L)}\widetilde{\sum}\prod\limits_{j\not\in
I_\alpha}\left(\displaystyle\int\limits_{0}^\infty
x_j^{m-n}d\,x_j\right)\prod\limits_{s=1}^k\lambda_s^{m-n}\\
&\times& e^{-m\sum\limits_{j\not\in I_\alpha}
x_j\tau_j-m\sum\limits_{s=1}^k\lambda_{s}\tau_{\alpha_s}}\triangle(X_\alpha),
\end{eqnarray*}
where
$$
\triangle(X_\alpha)=\triangle(X)\Big|_{x_{\alpha_j}=\lambda_j},\quad j=1,\dots,k
$$
and the sum is over all collection $I_\alpha:=\{\alpha_i\}_{i=1}^k$ such that $\alpha_i\ne\alpha_j$. Using
$\triangle(X)=\det[x_j^l]_{j,l=0}^{n-1}$ and rewriting this integral as a determinant and computing integrals
over $x_j$ we get
\begin{equation}\label{R1}
R_k^{(n)}(\lambda_1,\ldots,\lambda_k)
=\dfrac{n!}{Z_n\triangle(L)}\widetilde{\sum}e^{-m\sum\limits_{s=1}^k\lambda_{s}\tau_{\alpha_s}}
\det\left\{p_{j,l}^{\alpha}\right\}_{j,l=1}^{n},
\end{equation}
where
\begin{equation}\label{p_jl}
p_{j,l}^\alpha=\left\{
\begin{array}{ll}
\lambda_s^{m-n+l-1},& j=\alpha_s\,\,\hbox{for some}\, \,s,\\
\dfrac{(m-n+l-1)!}{(m\tau_j)^{m-n+l}},&\hbox{otherwise}.
\end{array}
\right.
\end{equation}
Note that
\begin{equation*}
\lambda_s^{m-n+l-1}=\dfrac{(m-n+l-1)!m}{2\pi
i}\displaystyle\oint\limits_\omega\dfrac{e^{mu\lambda_s}d\,u}{(mu)^{m-n+l}}
\end{equation*}
where $\omega$ is a closed contour encircling zero.
Hence, (\ref{R1}) can be rewritten as
\begin{eqnarray}\label{R2}\notag
R_k^{(n)}(\lambda_1,\ldots,\lambda_k) &=&\dfrac{n!m^k\prod\limits_{l=1}^n(m-n+l-1)!}{Z_n}\widetilde{\sum}
\left(\prod\limits_{s=0}^k\oint\dfrac{d\,u_s}{2\pi i}\right)\\
&\times&\dfrac{e^{-m\sum\limits_{s=1}^k\lambda_{s}\tau_{\alpha_s}+m\sum\limits_{s=1}^ku_s\lambda_s}}
{\prod\limits_{j\not\in I_\alpha} (m\tau_j)^m\prod\limits_{s=1}^k(mu_s)^m}
\dfrac{\det\left\{q_{j,l}^{\alpha}\right\}_{j,l=1}^{n}}{\triangle(L)},
\end{eqnarray}
where
\begin{equation}\label{q_jl}
q_{j,l}^\alpha=\left\{
\begin{array}{ll}
(mu_s)^{n-l},& j=\alpha_s\,\,\hbox{for some}\, \,s,\\
(m\tau_j)^{n-l},&\hbox{otherwise}.
\end{array}
\right.
\end{equation}
It is easy to see that
\[
\det\left\{q_{j,l}^{\alpha}\right\}_{j,l=1}^{n}=m^{n(n-1)/2}
\prod\limits_{s<t}(u_s-u_t)\prod\limits_{j<l,j,l\not\in I_\alpha}(\tau_j-\tau_l)
\prod\limits_{s=1}^k\prod\limits_{j\not\in I_\alpha} \varepsilon_{j,s}(u_s-\tau_j),
\]
where
\[
\varepsilon_{j,s}=\left\{
\begin{array}{ll}
+1,&\alpha_s<j,\\
-1,&\alpha_s>j.
\end{array}
\right.
\]
Using this we obtain
\begin{equation}\label{frac_det}
\dfrac{\det\left\{q_{j,l}^{\alpha}\right\}_{j,l=1}^{n}}{m^{n(n-1)/2}\triangle(L)}=\dfrac{
\prod\limits_{\alpha_s<\alpha_t}(u_s-u_t)}
{\prod\limits_{\alpha_s<\alpha_p}(\tau_{\alpha_s}-\tau_{\alpha_p})}\prod\limits_{s=1}^k\prod\limits_{j\not\in
I_\alpha} \dfrac{u_s-\tau_j}{\tau_{\alpha_s}-\tau_j}.
\end{equation}
Thus, according to the residue theorem (\ref{R2}) and (\ref{Z_n}) yield
\begin{eqnarray}\label{R_last}\notag
R_k^{(n)}(\lambda_1,\ldots,\lambda_k)
&=&m^k\prod\limits_{s=1}^k\displaystyle\oint\limits_{L}\dfrac{d\,t_s}{2\pi i}
\displaystyle\oint\limits_\omega\prod\limits_{s=1}^k\dfrac{d\,u_s}{2\pi i} \dfrac{e^{m\sum\limits_{s=1}^k
u_s\lambda_s-m\sum\limits_{s=1}^k\lambda_st_s} \prod\limits_{s=1}^kt_s^m}
{\prod\limits_{s=1}^ku_s^m}\\
&&\times(-1)^{\frac{k(k-1)}{2}}\dfrac{\prod\limits_{s<t}(u_s-u_t)\prod\limits_{s<p}(t_s-t_p)}
{\prod\limits_{s=1}^k\prod\limits_{p=1}^k
(u_s-t_p)}\prod\limits_{s=1}^k\prod\limits_{j=1}^n\dfrac{u_s-\tau_j}{t_s-\tau_j},
\end{eqnarray}
where $\tau_j=1/t_j^{(n)}$, $L$ is a closed contours encircling $\{\tau_j\}_{j=1}^n$ and $\omega$ is any
closed contour encircling $0$ and not intersecting $L$. Now the identity (see \cite{Po-Se:76}, Problem 7.3)
\[
(-1)^{\frac{k(k-1)}{2}}\dfrac{\prod\limits_{s<t}(u_s-u_t)\prod\limits_{s<p}(t_s-t_p)}
{\prod\limits_{s=1}^k\prod\limits_{p=1}^s (u_s-t_p)}=\det\left[\dfrac{1}{u_s-t_p}\right]_{p,s=1}^k,
\]
and (\ref{R_last}) yield (\ref{Det}) with (\ref{K}).
\end{proof}

\section{Proof of Theorem \ref{thm:1}.}
In this section we prove the universality conjecture (\ref{Un}) and (\ref{gp}) of the local bulk regime of Hermitian
random matrices (\ref{H}) in the conditions of Theorem \ref{thm:1} using (\ref{Det}) and
passing to the limit (\ref{c}) in (\ref{K}).

 Let us take some $\lambda_0$, $\rho(\lambda_0)>0$, where $\rho$ is defined in (\ref{rho}). Putting in formula
(\ref{K}) $\lambda=\lambda_0+\xi/n$ and $\mu=\lambda_0+\eta/n$, we get:
\begin{multline}\label{Ker1}
 \dfrac{1}{n} K_n(\lambda_0+\xi/n,\lambda_0+\eta/n)=\\
 -\dfrac{c_{m,n}}{4\pi^2}\oint\limits_L\oint\limits_\omega
\dfrac{\exp\{m((\lambda_0+\xi/n) u-(\lambda_0+\eta/n)t)\}}{u-t}\dfrac{t^{m}}
{u^{m}}\prod\limits_{j=1}^n\dfrac{u-\tau_j}{t-\tau_j}d\,u\,d\,t\\
= -\dfrac{c_{m,n}}{4\pi^2}\oint\limits_L\oint\limits_\omega \dfrac{\exp\{m(S_n(u,\lambda_0)-
S_n(t,\lambda_0))\} } {(u-t)}e^{(\xi u-\eta t)c_{m,n}}d\,u\,d\,t,
\end{multline}
where
\begin{equation}\label{S_n}
 S_n(z,\lambda_0)=\lambda_0 z-\ln z+
\dfrac{c_{m,n}^{-1}}{n}\sum\limits_{j=1}^n\ln (z-\tau_j)-S^*, \quad
c_{m,n}=\dfrac{m}{n},\quad\tau_j=1/t_j^{(n)},
\end{equation}
and $S^*$ is a constant which will be chosen later (see (\ref{S*})). Here $L$ is a
closed contour, encircling $\{\tau_j\}_{j=1}^n$ and $\omega$ is any closed contour encircling $0$ and not
intersecting $L$. Thus, (\ref{Ker1}) and (\ref{S_n}) yield
\begin{equation}\label{Ker}
\dfrac{1}{n}K_n(\lambda_0+\xi/n,\lambda_0+\eta/n)=- \dfrac{c_{m,n}}{4\pi^2}\oint\limits_L\oint\limits_\omega
\mathcal{F}_n(t,u)d\,u\,d\,t,
\end{equation}
where to simplify formulas below we denote
\begin{equation}\label{F_cal}
\mathcal{F}_n(t,u)=\dfrac{\exp\{m(S_n(u,\lambda_0)- S_n(t,\lambda_0))\} } {(u-t)}e^{(\xi u-\eta t)c_{m,n}}.
\end{equation}

We start from the following statement
\begin{proposition}
\label{p:rav_sh} Set
\begin{equation}\label{g_0,f_0}
g^{(0)}_n=\dfrac{1}{n}\sum\limits\dfrac{1}{\tau_j-z},\quad f^{(0)}=\int\dfrac{N^{(0)}(d\,t)}{\tau-z},\quad
\tau=1/t.
\end{equation}
Then we have
under conditions of Theorem \ref{thm:1}
\begin{equation}  \label{rav_sh}
\lim\limits_{n\to \infty}\mathbf{P}\{|g^{(0)}_n(z)-f^{(0)}(z)|>\varepsilon \}= 0
\end{equation}
uniformly in $z$ from compact set $K$ in the upper half-plane.
Moreover, the inverse assertion is true, i.e., if
(\ref{rav_sh}) is valid for some compact set $K$ in the upper half-plane, then we have (\ref{conpN0}).
\end{proposition}
The proof of the proposition is given, e.g., in \cite{TSh:08}.

  Let us take the disk $\sigma=\{z:\,|z_0-z|\le\varepsilon_1\}$ as the compact set $K$, where
\begin{equation}\label{z_0}
z_0=-c^{-1}f(\lambda_0-i0)-\dfrac{1-c^{-1}}{\lambda_0}
\end{equation}
with $f$ defined in (\ref{eqv_f}). Here we note that according to the result of \cite{Sil-Ch:95}
$f(\lambda_0-i0)$ exists and it is a continuous function for $\lambda_0>0$ and
\begin{equation}\label{z_0_rho}
\Im z_0=-c^{-1}\Im f(\lambda_0-i0)=c^{-1}\pi\rho(\lambda_0)>0.
\end{equation}
Taking into
account (\ref{rav_sh}), we get that for any sufficiently small $\delta>0$ and for any $\varepsilon>0$ there
exists $n_0$ such that for all $n>n_0$ the event
\begin{equation}\label{Om_e}
\Omega_\varepsilon=\{\max\limits_{z\in \sigma}|g^{(0)}_n(z)-f^{(0)}(z)|<\varepsilon\},
\end{equation}
satisfies the condition
\[
{\bf P}^{(\tau)}_n\{\Omega_\varepsilon^C\}\le \delta.
\]
 Recall that ${\bf P}^{(\tau)}_n$
and ${\bf E}^{(\tau)}_n$ correspond to the averaging with respect to $\Sigma_n$.

According to the determinant formulas (\ref{Det})-(\ref{K}),
we have to prove that for any $k\in\mathbb{N}$
\begin{multline}  \label{limr}
\lim\limits_{n\to \infty} \mathbf{E}^{(\tau)}_n\left\{\det\left\{\displaystyle
\frac{1}{n\rho_n(\lambda_0)}K_n\left(\lambda_0+\xi_i/ n\rho_n(\lambda_0),\lambda_0+
\xi_j/n\rho_n(\lambda_0) \right)\right\}_{i,j=1}^k\right\}=\\
\det\left\{S(\xi_i-\xi_j)\right\}_{i,j=1}^k,
\end{multline}
where $S$ is define in (\ref{S}).
Consider the expression
\begin{multline}\label{razn_int}
\mathbf{E}^{(\tau)}_n\left\{\det\left\{\displaystyle
\frac{1}{n\rho_n(\lambda_0)}K_n\left(\lambda_0+\xi_i/n\rho_n(\lambda_0),\lambda_0+
\xi_j/n\rho_n(\lambda_0) \right)\right\}_{i,j=1}^k\right.\\
\left. -\det\left\{S(\xi_i-\xi_j)\right\}_{i,j=1}^k\right\}.
\end{multline}
It is easy to see that to obtain (\ref{Un}) it suffices to prove that
\begin{equation}\label{Un2}
\begin{array}{cl}
\left|\dfrac{1}{n\rho_n(\lambda_0)} K_n\left(\lambda_0+\xi/n\rho_n(\lambda_0),\lambda_0+\eta/
n\rho_n(\lambda_0)\right)-\alpha(\xi,\eta)S(\xi-\eta)\right|\le
C\varepsilon, &\hbox{on} \,\,\Omega_{\varepsilon},\\
\left|\dfrac{1}{n\rho_n(\lambda_0)} K_n\left(\lambda_0+\xi/n\rho_n(\lambda_0),\lambda_0+\eta/
n\rho_n(\lambda_0)\right)\right|\le C,&\hbox{on} \,\,\Omega^C_{\varepsilon}.
\end{array}
\end{equation}
where $S$ and $K_n$ are defined in (\ref{S}) and (\ref{Ker}) respectively and $\alpha(\xi,\eta)$ is some
multiplier which vanishes during the calculation of the determinant in (\ref{Det}). Indeed, choose
sufficiently small $\varepsilon$ and $\delta$ and split $\mathbf{E}^{(\tau)}_n\{\ldots\} $ in
(\ref{razn_int}) into two parts, the integral over $\Omega_\varepsilon$ of (\ref{Om_e}) and the integral over
its complement. If we know (\ref{Un2}), then the first integral is small because the difference of the
determinants is small on $\Omega_\varepsilon$. In view of the Hadamard inequality (\cite{Co-Hi:53}, Section
I.5) and the second line of (\ref{Un2}), we have
\begin{multline}\label{ogr_det}
\left|\det \left\{\displaystyle\frac{1}{n\rho_n(\lambda_0)}K_n\left(\lambda_0+\xi_i/n\rho_n(\lambda_0),
 \lambda_0+\xi_j/n\rho_n(\lambda_0)\right)\right\}_{i,j=1}^{k}\right| \\
\le\prod\limits_{i=1}^k\left(\sum\limits_{j=1}^k\left|\displaystyle\frac{1}{n\rho_n(\lambda_0)}
K_n\left(\lambda_0+\xi_i/n\rho_n(\lambda_0), \lambda_0+\xi_j/n\rho_n(\lambda_0)\right)\right|^2\right)^{1/2}\le k^{k/2}C^k.
\end{multline}
Besides, $\det\{S(\xi_i-\xi_j)\}_{i,j=1}^k$, with $S$ of (\ref{S}), is bounded.
Thus, the integral over $\Omega_\varepsilon^C$ in (\ref{razn_int}) is bounded by $C\delta$ and hence we obtain
that (\ref{razn_int}) is bounded by $C\varepsilon_1$.
Thus, we are left to prove (\ref{Un2}).

 To do this we will choose the contour $L$ in (\ref{Ker}) as some $n$-dependent contour $L_n$.
 Consider the equation for $z$ with a real parameter $\lambda>0$
\begin{equation}\label{eqv_g_0_n}
 V(z):=\dfrac{1}{z}+c_{m,n}^{-1}g_n^{(0)}(z)=\lambda,
\end{equation}
where $g_{n}^{(0)}$ is defined in (\ref{g_0,f_0}). Equation (\ref{eqv_g_0_n}) can be written as a
polynomial equation of degree $(n+1)$ (if $\lambda>0$) and so it has $(n+1)$ roots. Consider $z\in
\mathbb{R}$. If $z\to \tau_j+0$, then $V(z)\to -\infty$, if $z\to \tau_j-0$, then $V(z)\to +\infty$, and also
if $z\to +0$, then $V(z)\to +\infty$, if $z\to -0$, then $V(z)\to -\infty$. Besides, $V(z)\to +0$, as
$z\to+\infty$ (since $c_{m,n}>1$) and $V(z)\to -0$, as $z\to -\infty$. Thus, the graph of $V(z)$ for
$z\in\mathbb{R}$ looks like in Fig.1.

\centerline{\includegraphics[width=6.0 in, height=4.0 in]{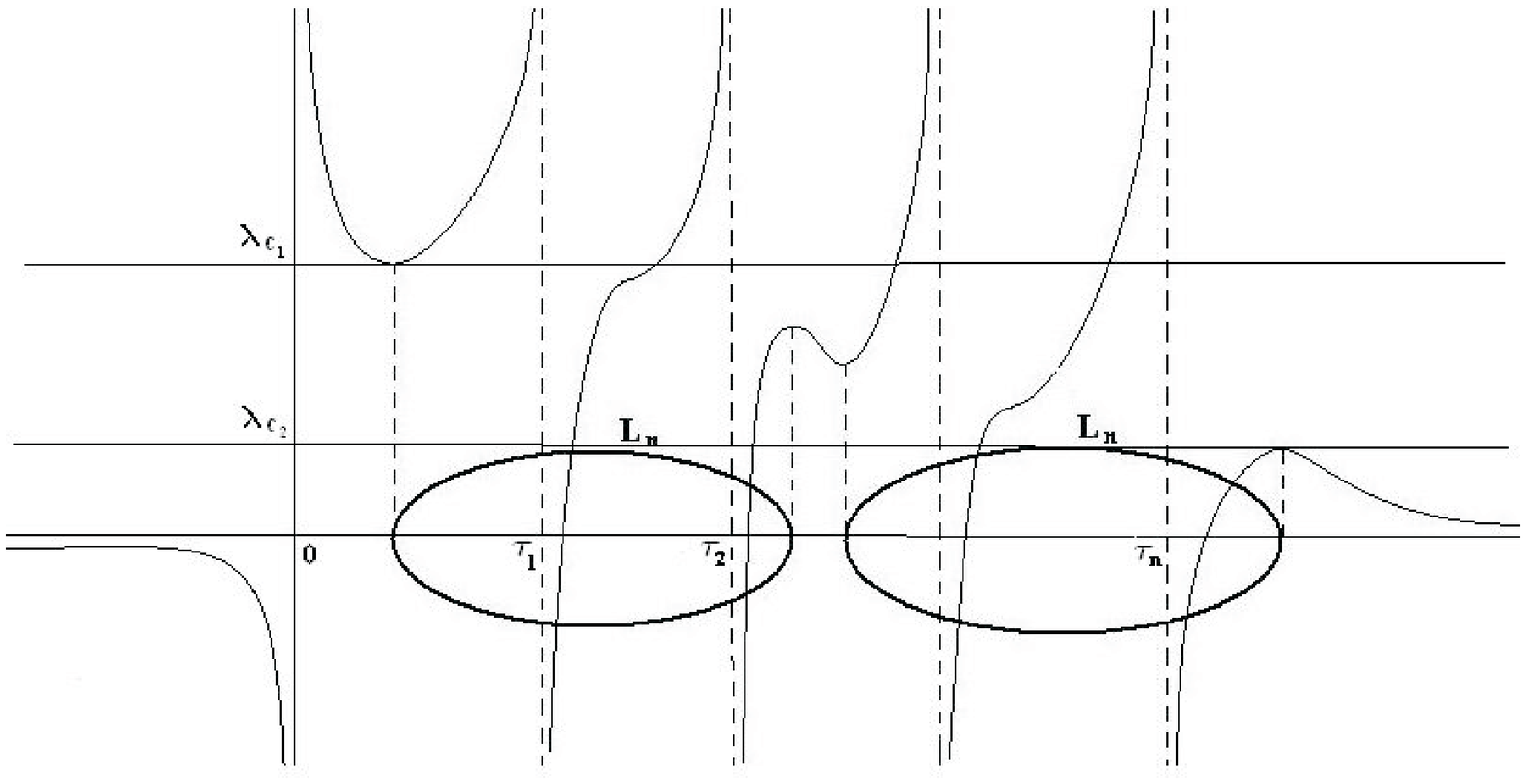}}

$n-1$ roots of (\ref{eqv_g_0_n}) are always real and lying between $\tau_j$-s. If $\lambda$ is big enough,
then all $n+1$ roots are distinct and real. Let $z_n(\lambda)$ be a root which tends to $0$, as
$\lambda\to\infty$. If $\lambda$ decreases, then at some $\lambda=\lambda_{c_1}$ two roots coincide
and for $\lambda\le\lambda_{c_1}$ the real root disappears and there appear two complex roots --
$z_n(\lambda)$ and $\overline{z_n(\lambda)}$. Indeed, it is easy to see, that since (\ref{eqv_g_0_n}) has not
more than $n+1$ roots, these coinciding roots are positive and smaller than any $\tau_j$. Thus
$\lambda_{c_1}>0$, since for $z^*=z_n(\lambda_{c_1})$ we have
\[
V(z^*)=\dfrac{1}{z_*}+\dfrac{c_{m,n}^{-1}}{n}\sum\limits_{j=1}^n\dfrac{1}{\tau_j-z_*}>0,
\]
because $z_*>0$ and $\tau_j>z_*$ for $j=1,..,n$. Then $z_n(\lambda)$ may be real again, than again complex,
and so on, however, as soon as $\lambda$ becomes less then some $\lambda_{c_2}>0$, $z_n(\lambda)$ becomes again
real (and it is real for any $0<\lambda<\lambda_{c_2}$). Moreover, $z_n(\lambda_{c_2})$ is bigger that every
$\tau_j$. Choose
\begin{multline}  \label{L_n}
L_n=\{z\in \mathbb{C}: z=z_n(\lambda):\,\Im z_n(\lambda)>0,\,\lambda>0\}\\
\cup\{z\in \mathbb{C}: z=\overline{z_n(\lambda)}:\,\Im z_n(\lambda)>0,\,\lambda>0\}\cup S^\prime,
\end{multline}
where $S^\prime$ is a set of points $z=z_n(\lambda)$ in which $z_n(\lambda)$ becomes real. Since (\ref{eqv_g_0_n})
always has exactly $n+1$ roots, it is clear that
the set of $\lambda$'s corresponding to $z_n(\lambda)\in L_n$ is $\bigcup\limits_{j=1}^k I_k$, where
$\{I_j\}_{j=1}^k$ are non intersecting segments. It is easy to see also that the contour $L_n$ is closed and
encircling $\{\tau_j\}_{j=1}^n$, and $L_n$ lies in the right half plane. In addition, we will use later that for real $z$
$V^\prime(z)$ can change the sign only in the points of $S^\prime$.

To prove (\ref{Un2}), let us prove first that
\begin{equation}\label{Un1}
\begin{array}{cl}
\left|\dfrac{1}{n} K_n\left(\lambda_0+\xi/n,\lambda_0+\eta/n\right)-
\alpha(\xi,\eta)\dfrac{\sin\left(\pi c_{m,n}y_n(\lambda_0)(\eta-\xi) \right)}{\pi
(\xi-\eta)}\right|\le
C\varepsilon, &\hbox{on} \,\,\Omega_{\varepsilon},\\
\left|\dfrac{1}{n} K_n\left(\lambda_0+\xi/n,\lambda_0+\eta/n\right)\right|\le C,&\hbox{on}
\,\,\Omega_{\varepsilon}^C,
\end{array}
\end{equation}
where $y_n(\lambda)=\Im z_n(\lambda)$, $K_n$ and $\Omega_\varepsilon$ are defined in (\ref{Ker}) and
(\ref{Om_e}), $z_n(\lambda)$ is a root of (\ref{eqv_g_0_n}) which tends to $+0$, as $\lambda\to\infty$, and
$\alpha(\xi,\eta)$ is some multiplier which vanishes during the computation of the determinant in
(\ref{Det}).

To do this we replace the integration over $\omega$ by the integration over $\omega_n$, where
\begin{equation}\label{om_n}
\omega_n=\{z\in \mathbb{C}: z=u(\varphi)= re^{i\varphi},\,\,\varphi\in [0;2\pi)\}
\end{equation}
with $r=|z_n(\lambda_0)|$. Consider the contour $\omega_{\delta}$ of Fig.2, where $\delta$ is small enough.
It will be shown below (see Lemmas \ref{l:min_L}, \ref{l:max_om}) that $L_n$ and $\omega_n$ has only two
points of intersection: $z_n(\lambda_0)$ and $\overline{z_n(\lambda_0)}$.

\centerline{\includegraphics[width=5.5 in, height=3.0 in]{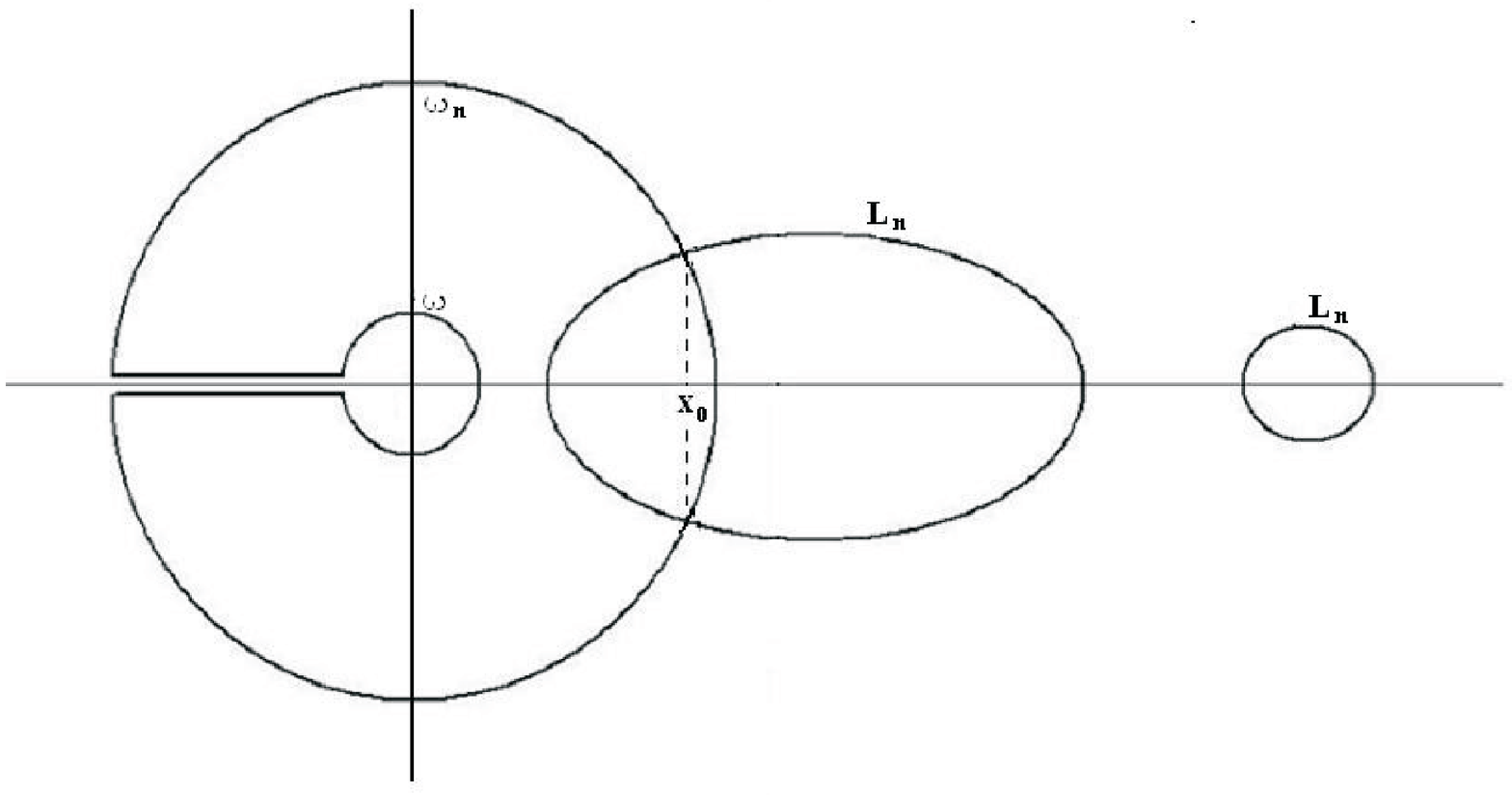}}

According to the residue theorem, we have
\begin{equation*}
\oint\limits_{\omega_{\delta}}\mathcal{F}_n(t,u)d\,u =\left\{
\begin{array}{cc}
2\pi i\cdot\exp\left\{(\xi-\eta)tc_{m,n}\right\},&t\,\,\hbox{is inside}\,\,\omega_\delta,\\
0,&t\,\,\hbox{is outside}\,\,\omega_\delta,\\
\pi i\cdot\exp\left\{(\xi-\eta)t c_{m,n}\right\} ,&t=z_n(\lambda_0),\overline{z_n(\lambda_0)},
\end{array}
\right.
\end{equation*}
where $\mathcal{F}_n$ is defined in (\ref{F_cal}). The sum of the integrals over the lines $\Im z=\pm\delta$
tends to $0$, as $\delta\to 0$, hence we get after the limit $\delta\to 0$
\begin{equation}
\label{Res}
\begin{array}{c}
\displaystyle{\oint\limits_{L_n}}\oint\limits_{\omega\,\cup\,\omega_n} \mathcal{F}_n(t,u)d\,u\,d\,t =2\pi
i\intd\limits_{z_n(\lambda_0)}^{\overline{z_n(\lambda_0)}}
\exp\left\{(\xi-\eta)t c_{m,n}\right\} d\,t\\
=4\pi c_{m,n}^{-1} e^{x_n(\lambda_0)(\xi-\eta)c_{m,n}}
\dfrac{\sin\left((\xi-\eta)y_n(\lambda_0)c_{m,n}\right)} {\xi-\eta}.
\end{array}
\end{equation}
Taking into account that
\[
\displaystyle{\oint\limits_{L_n}}\oint\limits_{\omega}
\mathcal{F}_n(t,u)d\,u\,d\,t=\displaystyle{\oint\limits_{L_n}}\oint\limits_{\omega_n}
\mathcal{F}_n(t,u)d\,u\,d\,t-\displaystyle{\oint\limits_{L_n}}\oint\limits_{\omega\,\cup\,\omega_n}
\mathcal{F}_n(t,u)d\,u\,d\,t
\]
and since the r.h.s. of (\ref{Res}) is bounded (see Lemma \ref{l:prexp} below), to prove (\ref{Un1})
it suffices to prove that
\begin{equation}\label{lim0}
\displaystyle{\oint\limits_{L_n}}\oint\limits_{\omega_n} \left|\mathcal{F}_n(t,u)\right|d\,u\,d\,t\le \left\{
\begin{array}{l}
C\varepsilon, \quad \hbox{on}\,\,\Omega_\varepsilon,\\
C, \quad \hbox{on}\,\,\Omega^C_\varepsilon,
\end{array}\right.
\end{equation}
where $\mathcal{F}_n(t,u)$ is defined in (\ref{F_cal}).

Now let us choose the constant in (\ref{S_n}) as
\begin{equation}\label{S*}
S^*=\Re\left(\lambda_0z_n(\lambda_0)-\ln z_n(\lambda_0)+ \dfrac{c_{m,n}^{-1}}{n}\sum\limits_{j=1}^n\ln
(z_n(\lambda_0)-\tau_j)\right)
\end{equation}
and study the behavior of the function $\Re S_n(z_n(\lambda),\lambda_0)$ of (\ref{S_n}) on the contours $L_n$
of (\ref{L_n}) and $\omega_n$ of (\ref{om_n}).

\begin{lemma}\label{l:min_L}
Let $z\in L_n$ of (\ref{L_n}) and $\Im z\ge 0$. Then for any set $\{\tau_j\}_{j=1}^n\in(\mathbb{R}^+)^n$ we
have
$$\Re S_n(z_n(\lambda),\lambda_0)\ge 0$$
and the equality holds only at $\lambda=\lambda_0$. Moreover, the function $\Re S_n(z_n(\lambda),\lambda_0)$
is strictly increasing for $\lambda>\lambda_0$ and strictly decreasing for $\lambda<\lambda_0$. The same is
valid for the lower part of $L_n$, i.e., $z\in L_n$, $\Im z< 0$.
\end{lemma}
\begin{lemma}\label{l:max_om}
Consider $u\in\omega_n$, $\Im u>0$. Then we have for any set $\{\tau_j\}_{j=1}^n\in(\mathbb{R}^+)^n$
$$\Re S_n(u,\lambda_0)\le -C|x-x_0|^2,$$
where  $x_0=\Re z_n(\lambda_0)$, $x=\Re u$.
The same is valid for the lower part of $\omega_n$.
\end{lemma}
These lemmas yield that for $t\in L_n$, $u\in\omega_n$ and for any set
$\{\tau_j\}_{j=1}^n\in(\mathbb{R}^+)^n$
\[
  \Re((S_n(u,\lambda_0)-S_n(t,\lambda_0)))\le 0,
\]
and the equality holds only if $u$ and $t$ are both equal to $z_n(\lambda_0)$ or $\overline{z_n(\lambda_0)}$.

To estimate $\mathcal{F}_n(t,u)$ of (\ref{F_cal}) we use also
\begin{lemma}\label{l:prexp}
There exists an $n$-independent $\delta>0$ such that for any $\lambda\in U_\delta(\lambda_0)$ uniformly in
$\{\tau_j\}_{j=1}^n\in(\mathbb{R}^+)^n$ the solution
$z_n(\lambda)$ of equation (\ref{eqv_g_0_n}) admits the following bounds
\begin{equation}
\label{preexp} 0<C_1<x_n(\lambda)<C_2,\quad 0<C_1<|z_n(\lambda)|<C_2.
\end{equation}
where $x_n(\lambda)=\Re z_n(\lambda)$ (here and below $U_\delta(a)=(a-\delta,a+\delta)$).
\end{lemma}
In particular, this means that for any set $\{\tau_j\}_{j=1}^n\in(\mathbb{R}^+)^n$ the radius $r$ of
$\omega_n$ of (\ref{om_n}) satisfies the inequality
\begin{equation}\label{ineqv_r}
0<C_1 <r<C_2.
\end{equation}
Now we split the integral in (\ref{lim0}) into two parts
\begin{equation}
\label{Int_sk}
\displaystyle{\oint\limits_{L_n}}\oint\limits_{\omega_n}
\left|\mathcal{F}_n(t,u)\right|d\,u\,d\,t \\
= \left(\oint\limits_{\omega_n}\int\limits_{L_n^A}+ \oint\limits_{\omega_n}\intd\limits_{L_n\backslash
L_n^A}\right) |\mathcal{F}_n(t,u)|d\,u\,d\,t
\end{equation}
where $L_n^A$ is the part of $L_n$ where $|x_n(\lambda)|\le A$, $x_n(\lambda)=\Re z_n(\lambda)$ and
$\mathcal{F}_n(t,u)$ is defined in (\ref{F_cal}).

The next lemma implies the bound for the length of $L_n$
\begin{lemma}\label{l:dl_kont}
Let $l(x)$ be the oriented length of $L_n$  between $x_0=x_n(\lambda_0)$ and $x$ ($l(x)\ge 0$ for $x>x_0$).
Then uniformly in $\{\tau_j\}_{j=1}^n\in(\mathbb{R}^+)^n$ $l(x)$ admits the bound:
$$|l(x_1)-l(x_2)|\le C_1|x_1-x_2|+C_2(|x_1|+|x_2|)+C_3.$$
Moreover, if $0<x_1<x_2<C$, $j=1,2$, then
$$l(x_2)-l(x_1)\le C\sqrt{x_2-x_1}.$$
\end{lemma}
To estimate the r.h.s. of (\ref{Int_sk}) we use
\begin{lemma}\label{l:int_okr}
There exists a constant $\delta>0$ such that for all $\{\tau_j\}_{j=1}^n\subset (\mathbb{R}^+)^n$ and any
$x\not\in U_\delta(x_0)$ we have
\begin{equation*}
\hbox{dist}\{z_n(x),\omega_n\}\ge\delta,
\end{equation*}
where $z_n(x)$ is $z_n(\lambda)$ which is expressed via $x_n(\lambda)$ (we can do it according to
(\ref{otr_pr})).

Moreover, if $L_{[x_1,x_2]}$ is a part of $L_n$ between lines $\Re z=x_1$ and $\Re z=x_2$, $x_0=\Re
z_n(\lambda_0)$ and $\displaystyle\mu^*=\min_{z\in L_{[x_1,x_2]}}\Re S_n(z,\lambda_0)$ (recall that $\mu^*\ge
0$). Then for any $\{\tau_j\}_{j=1}^n\in(\mathbb{R}^+)^n$
\begin{equation}\label{L_otr}
\int\limits_{L_{[x_1,x_2]}}\oint\limits_{\omega_n} \left|\mathcal{F}_n(t,u)\right|d\,u\,d\,t
 \le \left\{\begin{array}{ll}Ce^{-m\mu^*},&0<x_1<x_2\le x_0/2,\\
 Ce^{-m\mu^*}\sqrt{x_2-x_1}|\ln(x_2-x_1)|,& x_0/2<x_1<x_2<C.
 \end{array}\right.
\end{equation}
\end{lemma}
Thus, the first integral in (\ref{Int_sk}) is bounded by $C=C(A)$ (we split $L_n^A$ in two segments by the
point $x=x_0/2$ and take the sum of the bounds for this segments).

To prove that the second integral in (\ref{Int_sk}) is bounded we consider
the imaginary part of (\ref{eqv_g_0_n}). We get
\begin{equation}\label{cond}
\dfrac{c_{m,n}^{-1}}{n}\sum\limits_{j=1}^n\dfrac{1}{(x_n(\lambda)-\tau_j)^2+y_n^2(\lambda)}=
\dfrac{1}{x_n^2(\lambda)+y_n^2(\lambda)},
\end{equation}
where $x_n(\lambda)=\Re z_n(\lambda)$, $y_n(\lambda)=\Im z_n(\lambda)$. This and (\ref{preexp}) yield for
$|z_n(\lambda)|>A$
\begin{equation}\label{razn_log}
\begin{array}{c}
\dfrac{1}{n}\sum\limits_{j=1}^n(\ln |z_n(\lambda_0)-\tau_j|- \ln
|z_n(\lambda)-\tau_j|)=\dfrac{1}{n}\sum\limits_{j=1}^n\ln \left|1+\dfrac{z_n(\lambda_0)-z_n(\lambda)}
{z_n(\lambda)-\tau_j}\right|\\
\le \dfrac{1}{n}\sum\limits_{j=1}^n\dfrac{|z_n(\lambda_0)-z_n(\lambda)|}{|z_n(\lambda)-\tau_j|}
\le|z_n(\lambda_0)-z_n(\lambda)| \left(\dfrac{1}{n}\sum\limits_{j=1}^n\dfrac{1}
{|z_n(\lambda)-\tau_j|^2}\right)^{1/2}\\
\le \dfrac{c_{m,n}^{1/2}|z_n(\lambda_0)-z_n(\lambda)|}{|z_n(\lambda)|}\le C.
\end{array}
\end{equation}
Using (\ref{S*}), (\ref{razn_log}) and Lemma \ref{l:prexp}, we obtain for $z_n\in L_n\setminus L_n^A$
\begin{equation}\label{ots_S_vsp}
\begin{array}{c}
\Re S_n(z_n(\lambda),\lambda_0)=\lambda_0x_n(\lambda)- \ln|z_n(\lambda)|
+\dfrac{c_{m,n}^{-1}}{n}\sum\limits_{j=1}^n\ln |z_n(\lambda)-\tau_j|-S^*\\
\ge \lambda_0x_n(\lambda)- \dfrac{1}{2}\ln|z_n(\lambda)|^2+C.
\end{array}
\end{equation}
Moreover, it follows from (\ref{cond}) that there exists $j$ such that
\begin{equation*}
y_n^{-2}(\lambda)\ge\dfrac{1}{(x_n(\lambda)-\tau_j)^2+y^2_n(\lambda)}\ge \dfrac{c_{m,n}}{x_n^2(\lambda)
+y^2_n(\lambda)},
\end{equation*}
and hence
\begin{equation}\label{ineqv_y}
(c_{m,n}-1)y^2_n(\lambda)\le x_n^2(\lambda).
\end{equation}
Therefore, (\ref{ots_S_vsp}) yields
\begin{equation}\label{ots_S}
\Re S_n(z_n(\lambda),\lambda_0) \ge \lambda_0 x_n(\lambda)- C_1\ln|x_n(\lambda)|+C_2 \ge\dfrac{\lambda_0
x_n(\lambda)}{2}+C,
\end{equation}
if $|x_n(\lambda)|>A$, where $A\in \mathbb{N}$ is big enough, but independent of $\{\tau_j\}_{j=1}^n$ and
$n$. Besides, Lemma \ref{l:prexp} yields for sufficiently big $A$
\[
\dfrac{1}{|u-t|}\le \dfrac{1}{|\Re u -\Re t|}\le \dfrac{1}{A-r}\le C.
\]
This, Lemmas \ref{l:max_om}-\ref{l:prexp}, (\ref{ineqv_r}) and (\ref{ots_S}) imply (recall that $m/n=c_{m,n}$)
\[
\displaystyle{\int\limits_{L_n\setminus L_n^A}}\oint\limits_{\omega_n}
\left|\mathcal{F}_n(t,u)\right|d\,u\,d\,t
 \le C_1\displaystyle{\int\limits_A^\infty}e^{-nxC_2+C_3n}d\,
l(x),
\]
where $l(x)$ is defined in Lemma \ref{l:dl_kont} and $C_2>0$. According to Lemma \ref{l:dl_kont} we obtain for  sufficiently
big $A$ and $n$
\begin{eqnarray*}
\displaystyle{\int\limits_A^\infty}e^{-nC_2x+C_3n}d\, l(x)&=&\sum\limits_{k=A}^\infty
\displaystyle{\int\limits_k^{k+1}}e^{-nC_2x+nC_3}d\,
l(x)\\
 \le \sum\limits_{k=A}^\infty e^{-nkC_2+nC_3}
(l(k+1)-l(k))&\le& \sum\limits_{k=A}^\infty e^{-nkC_2+nC_3} (C_3+C_4k)<e^{-nd}
\end{eqnarray*}
for some $d>0$. Thus,
\begin{equation}\label{int_A}
\displaystyle{\int\limits_{L_n\setminus L_n^A}}\oint\limits_{\omega_n}
\left|\mathcal{F}_n(t,u)\right|d\,u\,d\,t
 \le e^{-nd},
\end{equation}
and we have proved the second line of (\ref{lim0}) (see (\ref{Int_sk}), Lemma \ref{l:int_okr}, and (\ref{int_A})).

Let us prove now the first line of (\ref{lim0}). Choose any sufficiently small $\delta>0$ and split the integral in (\ref{lim0}) into three parts
\begin{equation}
\label{Int_sk1} \displaystyle{\oint\limits_{L_n}}\oint\limits_{\omega_n}
\left|\mathcal{F}_n(t,u)\right|d\,u\,d\,t = \left(\oint\limits_{\omega_n}\intd\limits_{U_1}+
\oint\limits_{\omega_n}\int\limits_{L_n^A\backslash U_1}+
\oint\limits_{\omega_n}\intd\limits_{L_n\backslash L_n^A}\right) |\mathcal{F}_n(t,u)|d\,u\,d\,t
\end{equation}
where $U_1=\{z_n(\lambda),\overline{z_n(\lambda)}:\lambda\in U_\delta(\lambda_0)\}$ and $L_n^A$ is
defined in (\ref{Int_sk}). Use
\begin{lemma}\label{l:x'}
For $\{\tau_j\}_{j=1}^n\in \Omega_\varepsilon$ and for $\lambda\in U_\delta(\lambda_0)$ with sufficiently
small $\delta$ we have
\[
0<C_1<|x_n^\prime(\lambda)|<C_2.
\]
\end{lemma}
This lemma and Lemma \ref{l:int_okr} yield
\[
\displaystyle\oint\limits_{\omega_n}\intd\limits_{U_1}\left|\mathcal{F}_n(t,u)\right|d\,u\,d\,t\le
\displaystyle\oint\limits_{\omega_n}\intd\limits_{L_{\delta'}}\left|\mathcal{F}_n(t,u)\right|
\le C\sqrt{\delta}\ln\delta^{-1},
\]
where $U_1$ is defined in (\ref{Int_sk1}), $\delta'=O(\delta)$,  $L_{\delta'}$ is a part of $L_n$ between the lines $\Re z=x_n(\lambda_0)+\delta'$
and $\Re z=x_n(\lambda_0)-\delta'$ and $\delta$ is small enough.

To estimates the second integral in (\ref{Int_sk1}) we need more information about the behavior of $\Re
S_n(z,\lambda_0)$ on $L_n$. Consider the second derivative of $\Re S_n(z,\lambda_0)$. We have
\begin{lemma}\label{l:vt_pr}
There exist $C>0$ and $\delta_0>0$ such that for $\{\tau_j\}_{j=1}^n\in \Omega_\varepsilon$
$$\dfrac{d^2}{d\,\lambda^2}\Re(- S_n(z_n(\lambda),\lambda_0))<-C$$ for any
$\lambda\in U_{\delta_0}(\lambda_0)$.
\end{lemma}
According to Lemma \ref{l:vt_pr} and the equalities
\[
\Re
S_n(z_n(\lambda_0),\lambda_0)=0,\quad\dfrac{d}{d\,\lambda}S_n(z_n(\lambda),\lambda_0)\Big|_{\lambda=\lambda_0}=0,
\]
(see (\ref{S_n}), (\ref{S*}), and (\ref{S1})) we obtain for any $\delta<\delta_0$
\begin{equation*}
\Re(- S_n(z_n(\lambda),\lambda_0))<-C\dfrac{(\lambda-\lambda_0)^2}{2}, \,\,\lambda\in
U_\delta(\lambda_0).
\end{equation*}
Since the function $\Re\left(S_n(z_n(\lambda),\lambda_0)\right)$ is monotone for $\lambda\ne\lambda_0$ (see
Lemma \ref{l:min_L}), we get
\begin{equation*}
\Re(- S_n(z_n(\lambda),\lambda_0))<-C\dfrac{\delta^2}{2}, \quad \lambda\not\in
U_\delta(\lambda_0).
\end{equation*}
This and Lemmas \ref{l:dl_kont}-\ref{l:x'} yield
\[
\oint\limits_{\omega_n}\intd\limits_{L_n^A\setminus U_1}\left|\mathcal{F}_n(t,u)\right|d\,u\,d\,t\le C\exp\{-nC\delta^2/2\}.
\]
The bound for the last integral in (\ref{Int_sk1}) is obtained in (\ref{int_A}).
Therefore, we have proved (\ref{lim0}), and combining it with (\ref{Ker}) and (\ref{Res}), we get
\begin{equation}\label{Main}
\left| \dfrac{1}{n}K_n\left(\lambda_0+\xi/n,\lambda_0+\eta/n\right)-
e^{(\xi-\eta)x_n(\lambda_0)c_{m,n}} \dfrac{\sin\left((\xi-\eta)y_n(\lambda_0)c_{m,n}\right)}
{\pi(\xi-\eta)}\right|\le \left\{\begin{array}{l}
 C\varepsilon,\quad\hbox{on}\,\,\Omega_\varepsilon,\\
 C,\quad\hbox{on}\,\,\Omega_\varepsilon^C.
 \end{array}\right.
\end{equation}
Thus we proved the first line of (\ref{Un1}).
The second line of (\ref{Un1}) follows from (\ref{Main}),
Lemma~\ref{l:prexp} and the bound
\begin{equation*}
\dfrac{\sin\left((\xi-\eta) y_n(\lambda_0)c_{m,n}\right)} {\pi(\xi-\eta)}\le \pi^{-1} c_{m,n}
y_n(\lambda_0).
\end{equation*}
Therefore, (\ref{Un1}) is proved.

To prove (\ref{Un2}) we are left to prove that $y_n(\lambda_0)c_{m,n}=\rho_n(\lambda_0)+o(1)$, $n\to\infty$. Consider the limiting equation
\begin{equation}
\label{eqv_f_0} \dfrac{1}{z}+c^{-1}f^{(0)}(z)=\lambda,
\end{equation}
where $f^{(0)}$ is defined in (\ref{g_0,f_0})
and $\lambda\in \mathbb{R}$ is fixed.
\begin{lemma}\label{l:8}
There exists $\varepsilon^\prime$ such that for any $\lambda\in U_{\varepsilon^\prime}(\lambda_0)$ equation
(\ref{eqv_f_0}) has a unique solution $z(\lambda)$ in the upper half-plane $\Im z>0$,  moreover
$z(\lambda_0)=z_0$, where $z_0$ is defined in (\ref{z_0}).
 The solution $z(\lambda)$ is continuous in $\lambda \in U_{\varepsilon^\prime}(\lambda_0)$
 and $\Im z(\lambda_0)>0$.
\end{lemma}
Relation between $z(\lambda)$ and $z_n(\lambda)$ is given by
\begin{lemma}\label{l:9}
There exists $\delta$ such that for all $\lambda\in U_\delta(\lambda_0)$ and for sufficiently big $n$
\[
|z_n(\lambda)-z(\lambda)|<\varepsilon\quad\hbox{on}\,\, \Omega_\varepsilon.
\]
\end{lemma}

The determinant formulas (\ref{Det})-(\ref{K}) imply that
$
\rho_n(\lambda)=\frac{1}{n}{\bf E}^{(\tau)}_n\{K_n(\lambda,\lambda)\},
$
where $\rho_n$, $K_n$ is defined in (\ref{rhon}) and (\ref{K}). Then
\begin{multline*}
\rho_n(\lambda_0)-\rho(\lambda_0)\\
=\mathbf{E}_n^{(\tau)}\left\{(n^{-1}K_n(\lambda_0,\lambda_0)-\rho(\lambda_0))\mathbf{1}_{\Omega_\varepsilon}\right\}
+\mathbf{E}_n^{(\tau)}\left\{(n^{-1}K_n(\lambda_0,\lambda_0)-\rho(\lambda_0))\mathbf{1}_{\Omega_\varepsilon^C}\right\}.
\end{multline*}
According to the second line of (\ref{Un1}), the second term can be bounded by $C\delta$. According to Lemma \ref{l:9},
$|z_n(\lambda_0)- z(\lambda_0)|<\varepsilon_1$, where $z_n(\lambda_0)$ and $z(\lambda_0)$ are the solutions
of equation (\ref{eqv_g_0_n}) and (\ref{eqv_f_0}) for $\lambda=\lambda_0$. Therefore, using Lemma
\ref{l:prexp} and (\ref{z_0_rho}), we have
$$
\left|\pi^{-1}c y_n(\lambda_0)-\rho(\lambda_0)\right|<\varepsilon_1,
$$
where $y_n(\lambda)=\Im z_n(\lambda)$. This and the first line of (\ref{Main}) for $\xi=\eta=0$ yield on
$\Omega_\varepsilon$ for any $\varepsilon_1>0$
$$
|n^{-1}K_n(\lambda_0,\lambda_0)-\rho(\lambda_0)|<
|n^{-1}K_n(\lambda_0,\lambda_0)-\pi^{-1}cy_n(\lambda_0)|+|\pi^{-1}c y_n(\lambda_0)-\rho(\lambda_0)|\le
2\varepsilon_1
$$
as $m,n\to\infty$. Therefore, for any $\varepsilon_1>0$ there exists such $N$ that we have for any $n>N$
\[
|\rho_n(\lambda_0)-\rho(\lambda_0)|<C\varepsilon_1.
\]
Thus, $\rho_n(\lambda_0)>0$ for sufficiently big
$n$ and we can divide by it in (\ref{Un1}). Therefore, we obtain (\ref{Un2}) and thus (\ref{Un}).

Let us prove (\ref{gp}). It is well-known (see,e.g.,\cite{Me:91}) that
\begin{multline*}
E_{n}\left( \left[\lambda _{0}+a/\rho_{n}(\lambda_{0})n,\lambda_{0}+b/\rho_{n}
(\lambda_{0})n\right]\right)\\
=1+\sum\limits_{l=1}^\infty\dfrac{(-1)^l}{l!}
\displaystyle\int\limits_a^b\det\left\{n^{-1}K_n\left(\lambda_0+x_i/\rho_{n}(\lambda_{0})\,n,
\lambda_0+x_j/\rho_{n}(\lambda_{0})\,n\right)\right\}_{i,j=1}^l\prod\limits_{j=1}^ld\,x_j.
\end{multline*}
Thus, according to the dominant convergence theorem, (\ref{Un}) and (\ref{ogr_det}) yield (\ref{gp}).
Therefore, Theorem \ref{thm:1} is proved.

\subsection{Proofs of the lemmas}

\textbf{Proof of Lemma \ref{l:min_L}.}
  Differentiate $\Re S_n(z_n(\lambda),\lambda_0)$ with respect to $\lambda$.
Using (\ref{S_n}) and equation (\ref{eqv_g_0_n}), we obtain
\begin{equation}\label{S1}
\begin{array}{c}
\dfrac{d}{d\,\lambda}\Re S_n(z_n(\lambda),\lambda_0)= \Re
\left(z_n^\prime(\lambda)\left(\lambda_0-\dfrac{1}{z_n(\lambda)}-
c_{m,n}^{-1}g_n^{(0)}(z_n(\lambda))\right)\right)\\
=\Re\,z_n^\prime(\lambda)\left(\lambda_0-\lambda\right)= -x_n^\prime(\lambda)(\lambda-\lambda_0).
\end{array}
\end{equation}
Let us show now that
\begin{equation}\label{otr_pr}
x_n^\prime(\lambda)<0.
\end{equation}
Differentiating (\ref{eqv_g_0_n}) with respect to $\lambda$ we get
\begin{equation}\label{zn_pr}
z_n^\prime(\lambda)=\left(c_{m,n}^{-1}\dfrac{d}{d\,z}g_n^{(0)}(z_n(\lambda))-\dfrac{1}
{z_n^2(\lambda)}\right)^{-1}.
\end{equation}
It follows from the implicit function theorem that $L_n$ intersects the real axis at the points where
\[
c_{m,n}^{-1}\dfrac{d}{d\,x}g_n^{(0)}(x)-\dfrac{1}{x^2}=0.
\]
Since
\[
\dfrac{d}{d\,x}g_n^{(0)}(x)=\dfrac{1}{n}\sum\limits_{j=1}^n\dfrac{1}{(x-\tau_j)^2},
\]
the inequality  $c_{m,n}^{-1}\dfrac{d}{d\,x}g_n^{(0)}(x)-\dfrac{1}{x^2}>0$ holds near $\tau_j$. As we note
before, $V^\prime(x)$ change sign only at the points of $S^\prime$ in which $L_n$ intersects real axis. Thus, the function
$c_{m,n}^{-1}\dfrac{d}{d\,x}g_n^{(0)}(x)-\dfrac{1} {x^2}$ is always negative outside $L_n$. On the other
hand, $z_n(\lambda)=x_n(\lambda)$ for $x>0$ outside $L_n$ and in this case
\begin{equation*}
x^\prime_n(\lambda)=z^\prime_n(\lambda)= \left(c_{m,n}^{-1}\dfrac{d}{d\,z}g_n^{(0)}(z_n(\lambda))-\dfrac{1}
{z_n^2(\lambda)}\right)^{-1}<0.
\end{equation*}
Let now $\lambda\in\bigcup\limits_{j=1}^k I_j$, i.e., $z_n(\lambda)$ belongs to $L_n$. We get from
(\ref{zn_pr})
\begin{equation*}
\Re z_n^\prime(\lambda)=x_n^\prime(\lambda)=\dfrac{a_n(\lambda)} {a^2_n(\lambda)+b^2_n(\lambda)},\,\,\, \Im
z_n^\prime(\lambda)=y_n^\prime(\lambda)=-\dfrac{b_n(\lambda)} {a^2_n(\lambda)+b^2_n(\lambda)},
\end{equation*}
where
\begin{equation}
\label{ab}
\begin{array}{rl}
a_n(\lambda)&= \Re\left(c_{m,n}^{-1}\dfrac{d}{d\,z}g_n^{(0)}(z_n(\lambda))-\dfrac{1}
{z_n^2(\lambda)}\right) ,\\
b_n(\lambda)&= \Im\left(c_{m,n}^{-1}\dfrac{d}{d\,z}g_n^{(0)}(z_n(\lambda))-\dfrac{1} {z_n^2(\lambda)}\right).
\end{array}
\end{equation}
Using (\ref{cond}), we obtain
\begin{eqnarray}\label{a}\notag
a_n(\lambda)&=& \dfrac{c_{m,n}^{-1}}{n}\sum\limits_{j=1}^n\dfrac{(x_n(\lambda)-\tau_j)^2-y_n^2(\lambda)}
{|z_n(\lambda)-\tau_j|^4}-\dfrac{x^2_n(\lambda)-
y^2_n(\lambda)}{|z_n(\lambda)|^4}\\
&=&\dfrac{2y_n^2(\lambda)}{|z_n(\lambda)|^4}-\dfrac{c_{m,n}^{-1}}{n}\sum\limits_{j=1}^n\dfrac{2y^2_n(\lambda)}
{|z_n(\lambda)-\tau_j|^4}.
\end{eqnarray}
Besides, according (\ref{cond}) we have
\[
\dfrac{1}{n}\sum\limits_{j=1}^n\dfrac{1} {|z_n(\lambda)-\tau_j|^4}\ge
\left(\dfrac{1}{n}\sum\limits_{j=1}^n\dfrac{1} {|z_n(\lambda)-\tau_j|^2}\right)^2 =\dfrac{c_{m,n}^2}
{|z_\lambda|^4}.
\]
Then, since $c_{m,n}\to c>1$, as $m,n\to\infty$, (\ref{a}) yields
\begin{equation}\label{ineqv2}
a_n(\lambda)<-2y^2_n(\lambda)\dfrac{c_{m,n}-1}{|z_n(\lambda)|^4}<0
\end{equation}
Thus, in this case we also have
\begin{equation}\label{Rezpr}
x_n^\prime(\lambda)<0.
\end{equation}
According to (\ref{S1}), this means that the function $\Re S_n(z_n(\lambda),\lambda_0)$ is strictly
increasing
for $\lambda>\lambda_0$ and strictly decreasing for $\lambda<\lambda_0$, i.e. $\Re  S_n(z,\lambda_0)$ has a
minimum at $\lambda=\lambda_0$. Since $\Re S_n(z_n(\lambda_0),\lambda_0)=0$, $\Re
S_n(z_n(\lambda),\lambda_0)\ge 0$ and the equality holds only at $\lambda=\lambda_0$.

 Note that the lower part of $L_n$ differs from the upper one only
by the sign of $y_n(\lambda)$, hence $\Re S_n(z,\lambda_0)\ge 0$, $z\in L_n$ and the equality holds only at
$z=z_n(\lambda_0)$ and $z=\overline{z_n(\lambda_0)}$. $\quad\Box$
\medskip

\textbf{Proof of the Lemma \ref{l:max_om}.} $\Re S_n(z,\lambda_0)$ on $\omega_n$ of (\ref{om_n}) we can
rewrite as
\[
\Re S_n(u(\varphi),\lambda_0)=\lambda_0r\cos\varphi-\ln r+
\dfrac{c_{m,n}^{-1}}{2n}\sum\limits_{j=1}^n\ln(r^2-2r\tau_j\cos\varphi + \tau_j^2)- S^*.
\]
Differentiating this with respect to $\varphi\in (0,\pi)$ we
obtain
\begin{equation}
\label{Su1} \Re S_n(u(\varphi),\lambda_0)^\prime =r\sin\varphi\left(\dfrac{c_{m,n}^{-1}}{n}
\sum\limits_{j=1}^n\dfrac{\tau_j}{r^2-2r\tau_j\cos\varphi+ \tau_j^2}-\lambda_0\right).
\end{equation}
Set
\[
\varphi_0=\arg\, z_n(\lambda_0).
\]
It is easy to see (using (\ref{cond}) and (\ref{eqv_g_0_n})), that the expression in the brackets in
(\ref{Su1}) is zero at $\varphi=\varphi_0$, and, moreover, this expression is a strictly monotone decreasing
function of $\varphi$. Thus $\varphi=\varphi_0$ is a maximum of $\Re S_n(u(\varphi),\lambda_0)$. Similarly
the point $\varphi=-\varphi_0$ is a maximum for lower-half plane.

Besides, using that $r=|z_n(\lambda_0)|$, we have for $u\in\omega_n$
\[
\Re S_n(u,\lambda_0)=\lambda_0\widetilde{x}-\ln r+\dfrac{c_{m,n}^{-1}}{2n}\sum\limits_{j=1}^n \ln
(r^2-2\widetilde{x}\tau_j+\tau_j^2)-S^*,
\]
where $\widetilde{x}=\Re u$. Thus,
\[
\dfrac{d}{d\,\widetilde{x}}\Re S_n(u,\lambda_0)=\lambda_0-
\dfrac{c_{m,n}^{-1}}{n}\sum\limits_{j=1}^n\dfrac{\tau_j}{r^2-2\widetilde{x}\tau_j+\tau_j^2}.
\]
Differentiating with respect to $\widetilde{x}$ twice we get for $u\in\omega_n$, $\widetilde{x}=\Re u$
\[
\begin{array}{c}
\dfrac{d^2}{d\,\widetilde{x}^2}\Re S_n(u,\lambda_0)=-
\dfrac{2c_{m,n}^{-1}}{n}\sum\limits_{j=1}^n\dfrac{\tau_j^2}{(r^2-2\widetilde{x}\tau_j+\tau_j^2)^2} \le
-2c_{m,n}^{-1}\left(\dfrac{1}{n}\sum\limits_{j=1}^n\dfrac{\tau_j}{r^2-2\widetilde{x}\tau_j+
\tau_j^2}\right)^2\\
=-2c_{m,n} \left(\dfrac{d}{d\,\widetilde{x}}\Re S_n(u,\lambda_0)-\lambda_0\right)^2 \le
4c_{m,n}\lambda_0\dfrac{d}{d\,\widetilde{x}}\Re S_n(u,\lambda_0)-2c_{m,n}\lambda_0^2.
\end{array}
\]
Since we proved that $\widetilde{x}=x_0=\Re z_n(\lambda_0)$ is a maximum $\Re S_n(u,\lambda_0)$ on $\omega_n$, where $x_0=\Re
z_n(\lambda_0)$, and $\Re S_n(z_n(\lambda_0),\lambda_0)=0$, integrating from $x_0$ to $\widetilde{x}>x_0$ we
obtain
\[
\dfrac{d}{d\,\widetilde{x}}\Re S_n(u,\lambda_0)\le 4c_{m,n}\lambda_0\Re
S_n(u,\lambda_0)-2c_{m,n}\lambda_0^2(\widetilde{x}-x_0),
\]
where $\widetilde{x}=\Re u$.
Hence, the Gronuol lemma yields for $\widetilde{x}-x_0>0$
\[
\Re S_n(u,\lambda_0)\le \dfrac{\lambda_0}{2}(\widetilde{x}-x_0)+\dfrac{c_{m,n}^{-1}}{8}
\left(1-e^{4c_{m,n}\lambda_0(\widetilde{x}-x_0)}\right)\le -C(\widetilde{x}-x_0)^2,
\]
where $u\in\omega_n$, $\widetilde{x}=\Re u$. It is easy to see, that similar inequality we have for
$\widetilde{x}-x_0<0$ and therefore we get
\[
\Re S_n(u,\lambda_0)\le -C(\Re u-x_0)^2
\]
for any $u\in\omega_n$. Since the lower part of $\omega_n$ differs from the upper one only by the sign of
$\Im u$, the proof is complete. $\quad\Box$
\medskip

\textbf{ Proof of Lemma \ref{l:prexp}.} Equation (\ref{eqv_g_0_n}) has $(n+1)$ roots $z_1,\ldots,z_{n+1}$ for
$\lambda\ne 0$. According to the Viet theorem
\begin{equation*}
 z_1+z_2+\ldots+z_{n+1}=\dfrac{1-c_{m,n}^{-1}}{\lambda}+\sum\limits_{j=1}^n\tau_j.
\end{equation*}
Moreover, $n-1$ of these roots are always real and lying to the left of the corresponding $\tau_j$, and
hence their sum is less than $\sum\limits_{j=1}^n\tau_j$ (see Fig.1). Therefore, since we consider
$\lambda>0$, the sum of the last two roots are greater than $(1-c_{m,n}^{-1})/\lambda$.

If the last two roots of (\ref{eqv_g_0_n}) are real (i.e. $y_n(\lambda)=0$), then we obtain
\[
x^\prime_n(\lambda)=\left(\dfrac{c_{m,n}^{-1}}{n}
\sum\limits_{j=1}^n\dfrac{1}{(x_n(\lambda)-\tau_j)^2}-\dfrac{1} {x_n^2(\lambda)}\right)^{-1}.
\]
According to (\ref{otr_pr}), we get
\[
\begin{array}{c}
0<\dfrac{1} {x_n^2(\lambda)}-\dfrac{c_{m,n}^{-1}}{n} \sum\limits_{j=1}^n\dfrac{1}{(x_n(\lambda)-\tau_j)^2}<
\dfrac{1} {x_n^2(\lambda)}-c_{m,n}^{-1}\left(\dfrac{1}{n}
\sum\limits_{j=1}^n\dfrac{1}{x_n(\lambda)-\tau_j}\right)^2\\
=\dfrac{1} {x_n^2(\lambda)}- c_{m,n}\left(\dfrac{1} {x_n(\lambda)}-\lambda\right)^2.
\end{array}
\]
Therefore,
\[
\dfrac{1-c_{m,n}^{-1/2}}{\lambda}\le x_n(\lambda)\le \dfrac{1+c_{m,n}^{-1/2}}{\lambda}
\]
and for $\lambda>\delta>0$ such that $y_n(\lambda)=0$ the lemma is proved.

If the last two roots of (\ref{eqv_g_0_n}) are complex ($z_n(\lambda)$ and
$\overline{z_n(\lambda)}$) and $y_n(\lambda)\ne 0$, then
$$
\dfrac{1-c_{m,n}^{-1}}{\lambda}<z_n(\lambda)+\overline{z_n(\lambda)}=2x_n(\lambda).
$$
Moreover, we know from (\ref{eqv_g_0_n}) that
\[
\dfrac{1}{z_n(\lambda)}=\lambda+\dfrac{c_{m,n}^{-1}}{n} \sum\limits_{j=1}^n\dfrac{1}{z_n(\lambda)-\tau_j}.
\]
The real part of the l.h.s. of this equation is less than $\dfrac{1}{x_n(\lambda)}$. On the other hand, the
real part of the r.h.s. is greater than $\lambda-\dfrac{c_{m,n}^{-1}}{n}
\sum\limits_{j=1}^n\dfrac{|x_n(\lambda)-\tau_j|}{|z_n(\lambda)-\tau_j|^2}$. But for $z_n(\lambda)$ such that
$y_n(\lambda)\ne 0$, (\ref{cond}) yields
\[
\begin{array}{c}
\dfrac{1}{n}\sum\limits_{j=1}^n\dfrac{|x_n(\lambda)-\tau_j|}{|z_n(\lambda)-\tau_j|^2}\le \left(\dfrac{1}{n}
\sum\limits_{j=1}^n\dfrac{|x_n(\lambda)-\tau_j|^2}{|z_n(\lambda)-\tau_j|^2}\right)^{1/2} \left(\dfrac{1}{n}
\sum\limits_{j=1}^n\dfrac{1}{|z_n(\lambda)-\tau_j|^2}\right)^{1/2}\\
\le \dfrac{\sqrt{c_{m,n}}}{|z_n(\lambda)|} \le\dfrac{\sqrt{c_{m,n}}} {|x_n(\lambda)|}.
\end{array}
\]
Hence, for $\lambda>\delta>0$
\[
\dfrac{1+c_{m,n}^{-1/2}}{|x_n(\lambda)|}\ge \lambda \Rightarrow |x_n(\lambda)|\le
\dfrac{1+c_{m,n}^{-1/2}}{\lambda}<C,
\]
which complete the proof of the first inequality of (\ref{preexp}) for $y_n(\lambda)\ne 0$ (i.e.
$z_n(\lambda)\in L_n$).

The second inequality in (\ref{preexp}) for $\lambda: y_n(\lambda)\ne 0$ easily follows from the first
inequality and (\ref{ineqv_y}). The lemma is proved. $\quad\Box$
\medskip

\textbf{ Proof of Lemma \ref{l:dl_kont}.}
 It follows from (\ref{Rezpr}) and (\ref{a}) that one can express
$y_n(\lambda)$ via $x_n(\lambda)$ to obtain the "graph" $y_n(x)$ of the upper part of $L_n$. Denote
\begin{equation}\label{obozn}
\begin{array}{c}
y_n^2(x)=s(x),\quad x-\tau_j=\triangle_j,
\quad \sigma_0= \dfrac{1}{x^2+s}\\
\sigma_k= \dfrac{1}{n}\sumd\limits_{j=1}^n
 \dfrac{1}{(\triangle_j^2+s)^k},\quad
\sigma_{kl}= \dfrac{1}{n}\sumd\limits_{j=1}^n
\dfrac{\triangle^l_j}{(\triangle_j^2+s)^k},\,(k=\overline{1,3},\,l=1,2)\\
\end{array}
\end{equation}
Differentiating (\ref{cond}) with respect to $x$, we obtain the
equality
\[
s^\prime(x)\left(c_{m,n}^{-1}\sigma_2-\sigma_0^2\right)+ 2c_{m,n}^{-1}\sigma_{21}-2x\sigma_0^2=0.
\]
Note that
\[
\begin{array}{c}
x\sigma_0^2\le \sigma_0^{3/2},\quad \sigma_2\ge
\sigma_1^2=c_{m,n}^2\sigma_0^2\\ \\
\sigma_{21}\le \left(\dfrac{1}{n}\sumd\limits_{j=1}^n
\dfrac{\triangle^2_j}{(\triangle_j^2+s(x))^2}\right)^{1/2} \left(\dfrac{1}{n}\sumd\limits_{j=1}^n
\dfrac{1}{(\triangle_j^2+s(x))^2}\right)^{1/2}=
\sigma_{22}^{1/2}\sigma_2^{1/2}\le (c_{m,n}\sigma_0\sigma_2)^{1/2},\\
\end{array}
\]
thus
\begin{equation}\label{s1_1a}
\begin{array}{c}
(s^\prime(x))^2=\dfrac{4(x\sigma_0^2-c_{m,n}^{-1}\sigma_{21})^2}{(c_{m,n}^{-1}\sigma_2-\sigma_0^2)^2} \le
8\left(\dfrac{x^2\sigma_0^4}{(c_{m,n}^{-1}\sigma_2-\sigma_0^2)^2}+
\dfrac{c_{m,n}^{-2}\sigma_{21}^2}{(c_{m,n}^{-1}\sigma_2-\sigma_0^2)^2}\right)\\
\le 8\left(\dfrac{1}{(c_{m,n}-1)^2\sigma_0}+
\dfrac{c_{m,n}^{-1}\sigma_2}{(c_{m,n}-1)\sigma_0(c_{m,n}^{-1}\sigma_2-\sigma_0^2)}\right).
\end{array}
\end{equation}
Since
\[
\dfrac{c_{m,n}^{-1}\sigma_2}{c_{m,n}^{-1}\sigma_2-\sigma_0^2}=1+\dfrac{\sigma_0^2}{c_{m,n}^{-1}\sigma_2-\sigma_0^2}\le
1+\dfrac{1}{c_{m,n}-1},
\]
(\ref{s1_1a}) yields for $(x,y(x))\in L_n$
\begin{equation*}
(s^\prime(x))^2\le \dfrac{C}{\sigma_0}=C(x^2+y^2(x)).
\end{equation*}
Thus, according to Lemma \ref{l:prexp} we obtain for any $x$ such that $(x,y(x))\in L_n$ and $|x|<C$
\begin{equation}\label{s'}
|s^\prime(x)|\le C.
\end{equation}
Differentiating (\ref{cond}) with respect to $x$ twice, we have in our notations
\begin{multline*}
s^{\prime\prime}(x)(c_{m,n}^{-1}\sigma_2-\sigma_0^2)=2(s^\prime(x))^2(c_{m,n}^{-1}\sigma_3-\sigma_0^3)\\
+8s^\prime(x)(c_{m,n}^{-1}\sigma_{31}-x\sigma_0^3)+6(c_{m,n}^{-1}\sigma_2-\sigma_0^2)+
8s(x)(\sigma_0^3-c_{m,n}^{-1}\sigma_3).
\end{multline*}
Since $\sigma_3\ge (c_{m,n}\sigma_0)^3>c_{m,n}\sigma_0^3$, the r.h.s. is a quadratic trinomial of
$s^\prime(x)$ with a positive leading coefficient (without addition with $s(x)$). Hence, using
$c_{m,n}^{-1}\sigma_2- \sigma_0^2>0$ we get
\begin{equation}\label{s2_1}
\begin{array}{c}
s^{\prime\prime}(x)(c_{m,n}^{-1}\sigma_2-\sigma_0^2)\ge
8s(x)(\sigma_0^3-c_{m,n}^{-1}\sigma_3)-\dfrac{8(c_{m,n}^{-1}\sigma_{31}-x\sigma_0^3)^2}
{c_{m,n}^{-1}\sigma_3-\sigma_0^3}\\
\ge -8s(x)c_{m,n}^{-1}\sigma_3-\dfrac{16c_{m,n}^{-2}\sigma_{31}^2}
{c_{m,n}^{-1}\sigma_3-\sigma_0^3}-\dfrac{16x^2\sigma_0^6} {c_{m,n}^{-1}\sigma_3-\sigma_0^3}.
\end{array}
\end{equation}
Note that $s(x)\sigma_3\le \sigma_2$ and also
\[
\sigma_{31}^2=\left(\dfrac{1}{n}\sumd\limits_{j=1}^n \dfrac{\triangle_j}{(\triangle_j^2+s(x))^3}\right)^2\le
\sigma_{23}\sigma_3\le\sigma_2\sigma_3.
\]
Using these inequalities, we get from (\ref{s2_1})
\[
s^{\prime\prime}(x)(c_{m,n}^{-1}\sigma_2-\sigma_0^2) \ge
-8c_{m,n}^{-1}\sigma_2-\dfrac{16c_{m,n}^{-2}\sigma_2\sigma_3}
{c_{m,n}^{-1}\sigma_3-\sigma_0^3}-\dfrac{16\sigma_0^5} {c_{m,n}^{-1}\sigma_3-\sigma_0^3}.
\]
Thus, since
\[
\begin{array}{c}
c_{m,n}^{-1}\sigma_2-\sigma_0^2\ge (c_{m,n}-1)\sigma_0^2>0, \quad c_{m,n}^{-1}\sigma_3-\sigma_0^3\ge
(c_{m,n}^{2}-1)\sigma_0^3,\\ \\
\dfrac{c_{m,n}^{-1}\sigma_2}{c_{m,n}^{-1}\sigma_2-\sigma_0^2}=1+\dfrac{\sigma_0^2}{c_{m,n}^{-1}\sigma_2-\sigma_0^2}\le
\dfrac{c_{m,n}}{c_{m,n}-1},\,\,\\
\dfrac{c_{m,n}^{-1}\sigma_3}{c_{m,n}^{-1}\sigma_3-\sigma_0^3}=1+\dfrac{\sigma_0^3}{c_{m,n}^{-1}\sigma_3-\sigma_0^3}\le
\dfrac{c_{m,n}^{2}}{c_{m,n}^{2}-1},
\end{array}
\]
we obtain
\begin{equation}\label{s2}
s^{\prime\prime}(x)\ge -C.
\end{equation}

Let $y_n^\prime(x)=\frac{s^\prime(x)}{2\sqrt{s(x)}}>0$ when $x\in[x_1,x_2]$. Then we have
\begin{multline}\label{ots_l1}
l(x_2)-l(x_1)=\intd\limits_{x_1}^{x_2}\sqrt{1+(y_n^\prime(x))^2}d\,x=
\intd\limits_{x_1}^{x_2}\sqrt{1+\left(\dfrac{s^\prime(x)}{2\sqrt{s(x)}}\right)^2}d\,x
\\
\le\intd\limits_{x_1}^{x_2}\left(1+
\dfrac{s^\prime(x)}{2\sqrt{s(x)}}\right)d\,x=(x_2-x_1)+\sqrt{s_2}-\sqrt{s_1}
\le(x_2-x_1)+\sqrt{s_2-s_1},
\end{multline}
where $s_2=s(x_2)$, $s_1=s(x_1)$.
If we choose in (\ref{ots_l1}) $x_2$ being the maximum point of $s(x)$, then $s^\prime(x_2)=0$ and we can write
\[
s_1-s_2=\dfrac{s^{\prime\prime}(\zeta)(x_1-x_2)^2}{2},
\]
where $\zeta\in [x_1,x_2]$.
This and (\ref{s2}) imply
\[
0\le s_2-s_1\le C(x_1-x_2)^2.
\]
Hence, we get in view of (\ref{ots_l1})
\begin{equation}\label{ots_l2}
l(x_2)-l(x_1)\le C(x_2-x_1).
\end{equation}
We have similar inequality for $x_1>x_2$ and $y_n^\prime(x)<0$, $x\in [x_2,x_1]$ (we should consider
$l(x_1)-l(x_2)$). Take an arbitrary $x_1\in [x_0;x_2]$ and denote by $x_*$ and $x^*$ the nearest to $x_1$ and
$x_2$ extremal points of $s(x)$ in $[x_1,x_2]$ correspondingly. Then, splitting $[x_*,x_*] $ in the segments
of monotonicity of $y_n$ and using (\ref{ots_l1}), its analog for decreasing $y_n(x)$ and (\ref{ots_l2}) we
obtain
\begin{equation}\label{ots_l_kor}
\begin{array}{c}
l(x_2)-l(x_1)=l(x_*)-l(x_1)+l(x^*)-l(x_*)+l(x_2)-l(x^*)\\
\le
C(x^*-x_*)+(x_2-x^*)+\sqrt{|s_2-s^*|}+(x_*-x_1)+\sqrt{|s_*-s_1|},
\end{array}
\end{equation}
where
Let now $x_1$, $x_2$ be such that $|x_1|,|x_2|\le C$. Then (\ref{ots_l_kor}) and (\ref{s'}) yield
\begin{eqnarray*}
l(x_2)-l(x_1)\le C(x_2-x_1)+C_1\sqrt{x_2-x^*}+\sqrt{x_*-x_1}\le C\sqrt{x_2-x_1},
\end{eqnarray*}
and so the second statement of the lemma is proved.

Also it easy to see from (\ref{ots_l1}) or its analog for decreasing $y_n(x)$ that for any segment $[a,b]$
where $y_n(x)$ is monotone we have
\begin{equation}\label{ots_l3}
l(b)-l(a)\le |b-a|+|y_n(b)-y_n(a)|.
\end{equation}
Again take an arbitrary $x_1\in [x_0;x_2]$ and denote by $x_*$ and $x^*$ the nearest to $x_1$ and $x_2$
extremal points of $s(x)$ in $[x_1,x_2]$ correspondingly. Using (\ref{ots_l2}), its analog for decreasing
$y_n(x)$, and (\ref{ots_l3}) we obtain
\begin{eqnarray}\label{ots_l}\notag
l(x_2)-l(x_1)=l(x_*)-l(x_1)+l(x^*)-l(x_*)+l(x_2)-l(x^*)\\
\le C(x_2-x_0)+|y_n(x_2)-y_n(x^*)|+|y_n(x_1)-y_n(x_*)|.
\end{eqnarray}
Taking into account (\ref{ineqv_y}) it is easy to see that
\[
y_n(x)\le C_1|x|+C_2.
\]
This and (\ref{ots_l}) yield
\[
|l(x_2)-l(x_1)|\le C_1|x_2-x_1|+C_2(|x_1|+|x_2|)+C_3.
\]
$\quad\Box$
\medskip

\textbf{ Proof of Lemma \ref{l:int_okr}.} Note that since $r$ is bounded from both sides uniformly
in $\{\tau_j\}_{j=1}^n$ and $n$ (see (\ref{ineqv_r})), we have for
$|z|\le C_1/2$ (here $C_1$ is a constant from (\ref{ineqv_r}))
\begin{equation*}
|z-u(\varphi)|\ge r-C_1/2,
\end{equation*}
and the assertions of the lemma follow from Lemmas \ref{l:min_L}-\ref{l:max_om} and \ref{l:dl_kont}.
If $z\in L_n$, $|z|>C_1/2$ then similarly to (\ref{razn_log}) we get
\begin{eqnarray*}
\Re(S_n(z,\lambda_0)-S_n(u(\varphi),\lambda_0))=\lambda_0(z-u(\varphi))+\ln |u(\varphi)|-\ln |z|\\
+\dfrac{c_{m,n}^{-1}}{n}\sum\limits_{j=1}^n\ln |z_n(\lambda)-\tau_j|- \dfrac{c_{m,n}^{-1}}{n}\sum\limits_{j=1}^n\ln
|u(\varphi)-\tau_j|\le\dfrac{C_2|z_n(\lambda)-u(\varphi)|}{|\Im u(\varphi)|}
\end{eqnarray*}
This inequality, Lemmas \ref{l:min_L}, \ref{l:max_om} yield for  $z\in L_n$, $|z|>C_1/2$
and $u(\varphi)\in \omega_n$
\begin{eqnarray*}
C|\Re u(\varphi)-x_0|^2\le -\Re S_n(u(\varphi),\lambda_0)
\le \Re(S_n(z,\lambda_0)-S_n(u(\varphi),\lambda_0)) \le\dfrac{C_2|z-u(\varphi)|}{|\Im u(\varphi)|}.
\end{eqnarray*}
Hence, we obtain
\begin{equation}\label{in_cu_1}
|z-u(\varphi)|\ge C r|\Re u(\varphi)-x_0|^2\sin\varphi,\quad z\in L_n,\quad |z|>C_1/2,\quad
\varphi\in[0,\pi/2].
\end{equation}
Thus for any $u(\varphi)\in\omega_n$ there are no points of $L_n$ in the disk of radius $C r|\Re
u(\varphi)-x_0|^2\sin\varphi$ centered in $u(\varphi)$. Consider two curves (see Fig.3):
\begin{equation}\label{curv}
\begin{array}{c}
\widetilde{z}_{\pm}(\varphi)=r(1\pm C|\Re u(\varphi)-x_0|^2\sin\varphi)e^{i\varphi},\quad
\varphi\in[0,\pi/2) \\ \\
\hbox{or}\quad\left\{\begin{array}{l}
\widetilde{x}_{\pm}(\varphi)=r(\cos\varphi\pm C|\Re u(\varphi)-x_0|^2\sin\varphi\cos\varphi),\\
\widetilde{y}_{\pm}(\varphi)=r(\sin\varphi\pm C|\Re u(\varphi)-x_0|^2\sin^2\varphi),\\
\end{array}\right.
\end{array}
\end{equation}
where $x_0=\Re z_n(\lambda_0)$.

\centerline{\includegraphics[width=5.5 in, height=3.5 in]{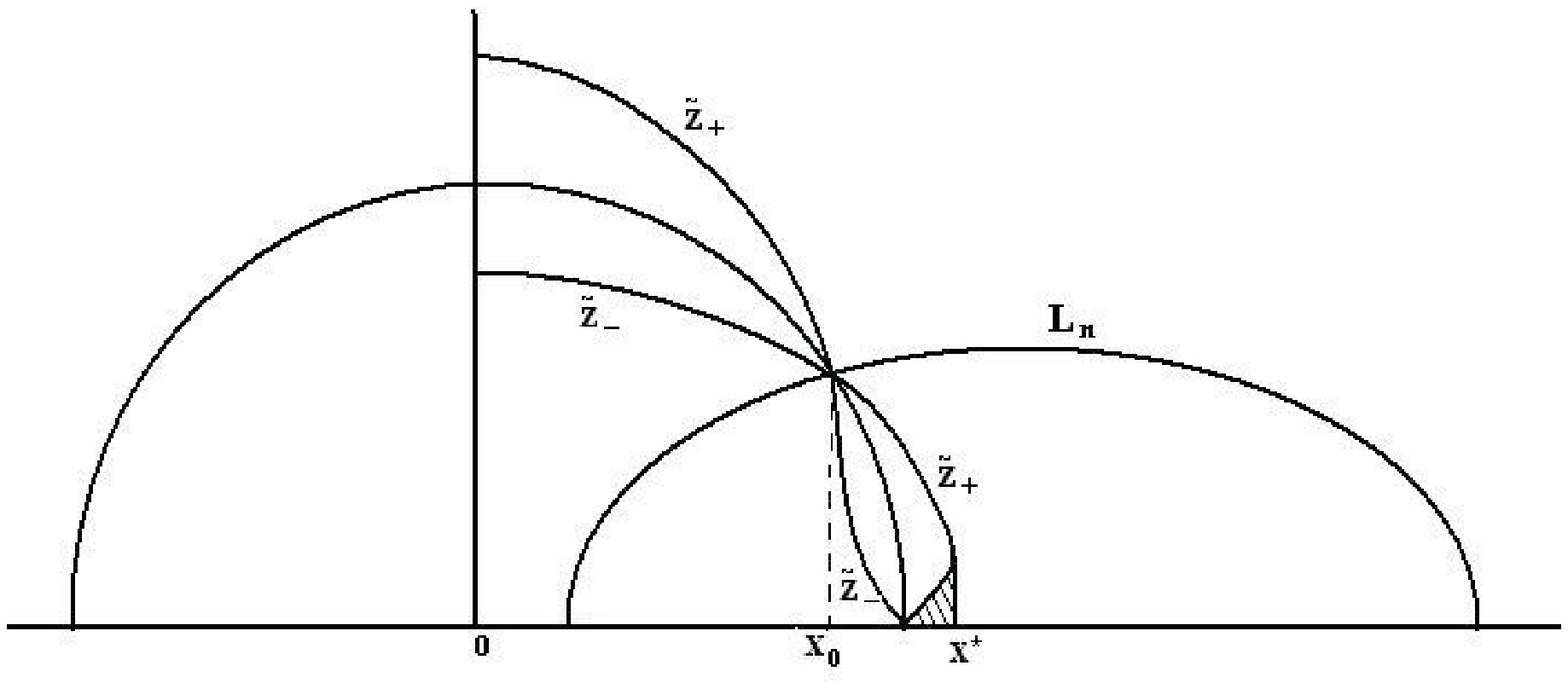}}

It is easy to see $\widetilde{x}_{+}^\prime(\varphi)<0$, $\widetilde{y}_{+}^\prime(\varphi)\ge 0$ when $\varphi$
is closed to $\pi$, and $\widetilde{x}_{+}^\prime(\varphi)>0$, $\widetilde{y}_{+}^\prime(\varphi)\ge 0$ when
$\varphi$ is closed to $\varphi_0=\arg z_n(\lambda_0)$. Let $\varphi^*$ be the nearest to $0$ root of
$\widetilde{x}_{+}^\prime(\varphi)$ on $[0,\pi/2]$, and $x^*=\widetilde{x}_{+}^\prime(\varphi^*)$. It is
evident that $x^*-r>C^*$, where $C^*$ is a constant depending only on the curve $\widetilde{z}_{+}(\varphi)$
and therefore bounded from below uniformly in $\{\tau_j\}_{j=1}^n$ and $n$. Moreover, there are no points of
$L_n$ in the domain $D^*$ bounded by the curve $\widetilde{z}_{+}(\varphi)$ and lines $\Im z=0$ and $\Re
z=x^*$. Indeed, since we know, that $L_n$ does not intersect $\widetilde{z}_{+}(\varphi)$, to come from
$z_n(\lambda_0)$ to any point of $D^*$, $z_n(\lambda)$ should intersect the line $\Re z=x^*$ twice (or more),
that contradict to (\ref{otr_pr}). Now taking into account that for $\lambda\to\infty$ $z_n(\lambda)$
is inside of $\omega_n$ and therefore for $x>x_0$ $z_n(x)$ is outside of $\omega_n$, we obtain the first
assertion of the lemma.

  Now, if $[x_1,x_2]\cap U_\delta(x_0)=\emptyset$, then (\ref{L_otr}) follows from Lemmas
\ref{l:min_L}-\ref{l:max_om} and from Lemma \ref{l:dl_kont}. If $[x_1,x_2]\subset U_\delta(x_0)$,
let us write
\[
\int\limits_{L_{[x_1,x_2]}}\oint\limits_{\omega_n} |\mathcal{F}_n(u,t)|d\,u\,d\,t \le
Ce^{-m\mu*}\int\limits_{L_{[x_1,x_2]}}\oint\limits_{\omega_n} \dfrac{|d\,u|\,|d\,t|} {|u-t|}\] Using that
$|t|\ge x_0/2$ for $t\in L_{[x_1,x_2]}$, $x_1>x_0/2$, we get
\[
\begin{array}{c}
\displaystyle{\oint\limits_{\omega_n} \dfrac{|d\,u|} {|u-t|}={r}\int\limits_0^{2\pi}\dfrac{d\,\varphi\,}
{\sqrt{r^2-2r|t|\cos\varphi+|t|^2}}= \int\limits_0^{2\pi}}\dfrac{\,d\,\varphi}
{\sqrt{(r-|t|)^2+4r|t|\sin^2\dfrac{\varphi}{2}}}\\
\le 2\sqrt{2}
\displaystyle{\int\limits_0^{\pi}}\dfrac{\,d\,\varphi}
{\left|r-|t|\right|+2\sqrt{r|t|}\sin\frac{\varphi}{2}}
\le C(1+\ln |r-|t||^{-1}),
\end{array}
\]
where $\varphi=\arg\,(u-t)$
Hence to prove the second statement of the Lemma \ref{l:int_okr} we should bound the
integral of $\ln |r-|t||^{-1}$.
To this end we use the curves $\widetilde{y}_\pm(x)$. Note that $\widetilde{y}_\pm^\prime(\varphi)>0$
and $\widetilde{x}_\pm^\prime(\varphi)>0$ for $\varphi\in U_{\delta_1}(\varphi_0)$. Hence, we can choose sufficiently
small $\delta$ (independent on $\{\tau_j\}_{j=1}^n$ and $n$) such that both curves $\widetilde{z}_\pm(\varphi)$
can be parameterized by $x\in U_\delta(x_0)$. Then according to the above argument we have
\begin{equation}\label{in_curv}
\left\{\begin{array}{lll}
y_n(x)\le \widetilde{y}_-(x),& x\le x_0,&x\in U_\delta(x_0),\\
y_n(x)\ge \widetilde{y}_+(x),& x\ge x_0,&x\in U_\delta(x_0).
\end{array}\right.
\end{equation}
Note that since $\cos\varphi,\sin\varphi>C_0>0$ for $\widetilde{x}_\pm(\varphi)\in U_\delta(x_0)$, we get
\begin{eqnarray*}
|\widetilde{x}_\pm(\varphi)-r\cos\varphi|&\ge& C|r\cos\varphi-x_0|^2
=C|r\cos\varphi-\widetilde{x}_\pm(\varphi)+
\widetilde{x}_\pm(\varphi)-x_0|^2\\
&\ge& C(|\widetilde{x}_\pm(\varphi)-x_0|^2-2|\widetilde{x}_\pm(\varphi)-x_0|\cdot|r\cos\varphi-\widetilde{x}_\pm(\varphi)|).
\end{eqnarray*}
Therefore,
\begin{equation}
\label{in1}
|\widetilde{z}_{\pm}(\varphi)-u(\varphi)|\ge|\widetilde{x}_{\pm}(\varphi)-r\cos\varphi|\ge C|\widetilde{x}_\pm(\varphi)-x_0|^2.
\end{equation}
Thus, for $x_0-\delta\le x_1<x_2\le x_0$, using (\ref{in_curv})-(\ref{in1}) we obtain
\begin{equation}\label{in2}
\begin{array}{c}
I(x_1,x_2):=-\displaystyle\int\limits_{x_1}^{x_2}\ln\left(r-\sqrt{x^2+y_n^2(x)}\right)d\,l(x)
\le-\displaystyle\int\limits_{x_1}^{x_2}\ln\left(r-\sqrt{x^2+\widetilde{y}_-^2(x)}\right)d\,l(x)\\
=-\displaystyle\int\limits_{x_1}^{x_2}\ln\left|u(\varphi_-(x))-\widetilde{z}_-(x)\right|d\,l(x)
\le -\int\limits_{x_1}^{x_2}\ln\left(C|x-x_0|^2\right)d\,l(x).
\end{array}
\end{equation}
Besides, Lemma \ref{l:dl_kont} implies that for any $x_1<x_2\le x_0$ we have
\begin{equation}\label{b_int_g}
\begin{array}{c}
-\displaystyle\int\limits_{x_1}^{x_2}\ln|x-x_0|d\,l(x)=
- (l(x_1)-l(x_2))\ln |x_1-x_0|
-\displaystyle\int\limits_{x_1}^{x_2}\dfrac{l(x)-l(x_2)}{x_0-x}d\,x
\\\le C\sqrt{x_2-x_1}|\ln (x_2-x_1)|+C\displaystyle\int\limits_{x_1}^{x_2}\dfrac{\sqrt{x_2-x}}{x_2-x}d\,x
\le C\sqrt{x_2-x_1}|\ln (x_2-x_1)|.
\end{array}
\end{equation}
For $x_0\le x_1<x_2\le x_0+\delta$ the proof is similar, we need only use that $y_n(x)\ge \widetilde{y}_+(x)$
in (\ref{in2}).
Now it is straightforward to obtain  the statement of
Lemma \ref{l:int_okr} for any $x_1,x_2>x_0/2$ (if it is necessary, we split $[x_1,x_2]$ on the segments by
the points $x_0-\delta$ and $x_0+\delta$ and then take the sum of the bounds for these segments).
$\quad\Box$
\medskip

\textbf{Proof of Lemma \ref{l:8}.} The existence of $f(\lambda_0\pm i0)$ is proved in \cite{Sil-Ch:95}. Using (\ref{eqv_f}) we obtain
\begin{equation}\label{eqv_l0}
\dfrac{1}{z(\lambda_0)}+c^{-1}f^{(0)}(z(\lambda_0))=\lambda_0,
\end{equation}
where $z(\lambda_0)$ is defined in (\ref{z_0}).
Hence, the solution exists if $\lambda=\lambda_0$.
Besides, for $x(\lambda_0)=\Re z(\lambda_0)$, $y(\lambda_0)=\Re z(\lambda_0)$ we have
\begin{multline}\label{pr_z}
\Re\left(\dfrac{1}{z}+c^{-1}f^{(0)}(z)\right)^\prime_z\Bigg|_{z=z(\lambda_0)}\\
=c^{-1}\intd\dfrac{((x(\lambda_0)-\tau)^2-y^2(\lambda_0))\,N^{(0)}(d\,t)}
{((x(\lambda_0)-\tau)^2+y^2(\lambda_0))^2} -\dfrac{(x^2(\lambda_0)-y^2(\lambda_0))}
{(x^2(\lambda_0)+y^2(\lambda_0))^2},
\end{multline}
where $\tau=1/t$.
Since $y(\lambda_0)\ne 0$, considering the l.h.s. of (\ref{eqv_l0}) we get
\[
\dfrac{1}{x^2(\lambda_0)+y^2(\lambda_0)}= c^{-1}\intd\dfrac{N^{(0)}(d\,t)}
{(x(\lambda_0)-\tau)^2+y^2(\lambda_0)}.
\]
Thus, we can rewrite (\ref{pr_z}) as
\begin{multline}\label{pr_z1}
\Re\left(\dfrac{1}{z}+c^{-1}f^{(0)}(z)\right)^\prime_z\Bigg|_{z=z(\lambda_0)}\\=
-\left(\intd\dfrac{2c^{-1}y^2(\lambda_0)N^{(0)}(d\,t)} {((x(\lambda_0)-\tau)^2+y^2(\lambda_0))^2}
-\dfrac{2y^2(\lambda_0)}{(x^2(\lambda_0)+y^2(\lambda_0))^2}\right),
\end{multline}
where $\tau=1/t$.
But we know that
\[
\intd\dfrac{N^{(0)}(d\,t)} {((x(\lambda_0)-\tau)^2+y^2(\lambda_0))^2}\ge \left(\intd\dfrac{N^{(0)}(d\,t)}
{(x(\lambda_0)-\tau)^2+y^2(\lambda_0)}\right)^2=\dfrac{c^2} {(x^2(\lambda_0)+y^2(\lambda_0))^2}.
\]
Therefore, since $c>1$, (\ref{pr_z1}) yields
\[
\Re\left(\dfrac{1}{z}+c^{-1}f^{(0)}(z)\right)^\prime_z\Bigg|_{z=z(\lambda_0)}< -\dfrac{2y^2(\lambda_0)(c-1)}
{(x^2(\lambda_0)+y^2(\lambda_0))^2}<0.
\]
Hence, according to the implicit function theorem, equation (\ref{eqv_f_0})
has a unique solution in the upper half-plane $\Im z>0$ if $\lambda
\in U_\varepsilon(\lambda_0)$, moreover the solution is continuous
in $\lambda$ in this neighborhood.
$\quad\Box$
\medskip

\textbf{Proof of the Lemma \ref{l:vt_pr}.}
We get from (\ref{S1})
\begin{equation}
\label{S2} \dfrac{d^2}{d\,\lambda^2}\Re(-
S_n(z_n(\lambda),\lambda_0))=\dfrac{d}{d\,\lambda}\left(x_n^\prime(\lambda)
(\lambda-\lambda_0)\right)=x_n^\prime(\lambda)+ x_n^{\prime\prime}(\lambda)(\lambda-\lambda_0).
\end{equation}
Hence, to prove the lemma it is sufficient to show that $x_n^{\prime\prime}(\lambda)$ is bounded uniformly in
$n$ and that $x_n^{\prime}(\lambda)$ is bounded from below by a positive constant uniformly in $n$ in some
sufficiently small neighborhood $U_\delta(\lambda_0)$.

 Show that $-x_n^{\prime}(\lambda)\ge C_1$ for all $\lambda\in
U_\delta(\lambda_0)$, where $C_1$ is a positive $n$-independent
constant. We have from (\ref{zn_pr})
\[
-\Re z_n^\prime(\lambda)=-x_n^\prime(\lambda)=\dfrac{-a_n(\lambda)} {a^2_n(\lambda)+b^2_n(\lambda)},
\]
where $a_n,\,b_n$ are defined in (\ref{ab}). Let us prove that $a_n(\lambda)$ and $b_n(\lambda)$ are bounded
uniformly in $n$ and that $a_n(\lambda)$ is bounded from below by a positive constant uniformly in $n$ for
all $\lambda\in U_\delta(\lambda_0)$ (evidently, it will be sufficient).

  Note that for $z_n(\lambda)\in L_n$
\begin{equation*}
\begin{array}{c}
|b_n(\lambda)|=\left|\dfrac{2y_n(\lambda) x_n(\lambda)} {|z_n(\lambda)|^4}-\dfrac{c_{m,n}^{-1}}{n}
\sum\limits_{j=1}^n\dfrac{2y_n(\lambda)(x_n(\lambda)-\tau_j)} {|z_n(\lambda)-\tau_j|^4} \right| \le
\dfrac{2|y_n(\lambda)||x_n(\lambda)|}
{|z_n(\lambda)|^4}\\
+ \dfrac{c_{m,n}^{-1}}{n} \sum\limits_{j=1}^n\dfrac{2|y_n(\lambda)||x_n(\lambda)-\tau_j|}
{|z_n(\lambda)-\tau_j|^4} \le\dfrac{1}{|z_n(\lambda)|^2}+ \dfrac{c_{m,n}^{-1}}{n}
\sum\limits_{j=1}^n\dfrac{1} {|z_n(\lambda)-\tau_j|^2} =\dfrac{2}{|z_\lambda|^2}
\end{array}
\end{equation*}
Also we can see from (\ref{a}) that
\begin{equation*}
|a_n(\lambda)|\le\left|\dfrac{2y_n^2(\lambda)} {|z_n(\lambda)|^4}+\dfrac{c_{m,n}^{-1}}{n}
\sum\limits_{j=1}^n\dfrac{2y^2_n(\lambda)} {|z_n(\lambda)-\tau_j|^4} \right| \le\dfrac{4} {|z_n(\lambda)|^2},
\end{equation*}
and besides (\ref{ineqv2}) yields
\[
-a_n(\lambda)>2y^2_n(\lambda)\dfrac{c_{m,n}-1}{|z_n(\lambda)|^4}.
\]

Using Lemma \ref{l:9}, continuity of $z(\lambda)$, and since
$\Im z(\lambda_0)=\pi\rho(\lambda_0)>0$, we have starting from some $n$
\begin{equation}\label{ogr}
|x_n(\lambda)|<C_1,\quad |y_n(\lambda)|<C_1, \quad |y_n(\lambda)|>C_2
\end{equation}
for all $\lambda\in U_\delta(\lambda_0)$, where $n$-independent $\delta$
is small enough.

 This fact yields that $-x_n^{\prime}(\lambda)\ge C$ for all
$\lambda\in U_\delta(\lambda_0)$, where $C$ is a positive $n$-independent constant, and also that
$x_n^{\prime\prime}$ is bounded uniformly and hence the second terms in (\ref{S2}) is of order $\delta$. The
lemma is proved. $\quad\Box$
\medskip

\textbf{ Proof of Lemma \ref{l:9}.}
We use Lemma \ref{l:8}. Consider the
solution $z(\lambda)$ of the limiting equation (\ref{eqv_f_0}) in
some neighborhood of $\lambda_0$. Since $\lambda_0\in \hbox{supp}\,N$,
$\Im z(\lambda_0)=A>0$. Taking into account the continuity of $z(\lambda)$
near $\lambda_0$, we can take a sufficiently small neighborhood
$U_{\delta_1}(\lambda_0)$ such that
\[
|z(\lambda)-z(\lambda_0)|<\varepsilon/2,\,\,\lambda\in
U_{\delta_1}(\lambda_0).
\]
Consider the set of the functions $f_\lambda(z)=\dfrac{1}{z}+c^{-1}f^{(0)}(z)-\lambda$ and the function
$$\phi(z)=c_{m,n}^{-1}g_n^{(0)}(z)-c^{-1}f^{(0)}(z),$$
where $f^{(0)},\,g_n^{(0)}$ are defined in (\ref{g_0,f_0}), and set $\omega=\{z:
|z-z(\lambda_0)|\le\varepsilon\}$. Let us show that for any $\lambda\in U_{\delta_1}(\lambda_0)$ and for any
$z\in\omega$
\begin{equation}
\label{Otgr_f}
|f_\lambda(z)|>c_0,
\end{equation}
where $c_0$ does not depend on $\lambda$. Assume the opposite and choose a sequence $\{\lambda_k\}_{k\ge 1}$,
$\lambda_k\in U_{\delta_1}(\lambda_0)$ such that $|f_{\lambda_k}(z_k)|\to 0$, as $k\to \infty$. There exists
a subsequence $\{\lambda_{k_l}\}$, converging to some $\lambda\in U_{\delta_1}(\lambda_0)$ such that the
subsequence $\{z_{k_l}\}$ converges to $z\in\partial\omega$. For these $\lambda$ and $z$ $f_{\lambda}(z)=0$.
But equation $f_{\lambda}(z)=0$ has in the upper half-plane only one root $z(\lambda)$, which is inside of
the circle of the radius $\varepsilon/2$ and with the center $z(\lambda_0)$. This contradiction proves
(\ref{Otgr_f}). Since $|g^{(0)}_n(z)- f^{(0)}(z)|\le \varepsilon$ on $\Omega_\varepsilon$ uniformly on
$\omega$ (see (\ref{Om_e}) and $c_{m,n}\to c$, we have starting from some~$n$
\begin{equation}
\label{Ogr_phi}
|\phi(z)|<c_0, \,\, z\in \partial\omega.
\end{equation}
Comparing (\ref{Otgr_f}) and (\ref{Ogr_phi}), we obtain that
starting from some $n$
\[
|f_\lambda(z)|>|\phi(z)|, \,\,z\in\partial\omega,\,\,
\forall\lambda\in U_{\delta_1}(\lambda_0).
\]
Since both functions are analytic, the Rouchet theorem implies that $f_\lambda(z)$ and
$f_\lambda(z)+\phi(z)=\dfrac{1}{z}+ c_{m,n}^{-1}g_n^{(0)}(z)-\lambda$ have the same number of zeros in
$\omega$. Since $f_\lambda(z)$ has only one zero in $\omega$, we conclude that $z_n(\lambda)$ belongs to
$\omega$, and so the lemma is proved. $\quad\Box$
\medskip

\textbf{ Proof of Lemma \ref{l:x'}.}
We have from (\ref{zn_pr})
\[
|z_n^\prime(\lambda)|^2=\dfrac{1}{a_n^2(\lambda)+b_n^2(\lambda)},
\]
where $a_n,\,b_n$ are defined in (\ref{ab}). Thus, using (\ref{ineqv2}), we obtain
\[
|z_n^\prime(\lambda)|\le\dfrac{1}{|a_n(\lambda)|}\le \dfrac{(x_n^2(\lambda)+
y^2_n(\lambda))^2}{2y^2_n(\lambda)(c_{m,n}-1)}.
\]
This and (\ref{ogr}) prove that $|x_n^\prime(\lambda)|<C$ for $\lambda\in U_\delta(\lambda_0)$. The
inequality $|x_n^\prime(\lambda)|>C$ was proved in Lemma \ref{l:vt_pr}. $\quad\Box$


\medskip


\begin{thebibliography}{99}
\bibitem{Pa-Sh:97} {\it L. Pastur, M. Shcherbina}, Universality of the local
eigenvalue statistics for a class of unitary invariant random
matrix ensembles. - J. Stat. Phys.(1997), 86, p.109-147

\bibitem{De-Co:99} {\it P. Deift, T. Kriecherbauer, K. McLaughlin, S. Venakides, X. Zhou},
Uniform asymptotics for polynomials orthogonal with respect to
varying exponential weights and applications to universality
questions in random matrix theory. - Commun. Pure Appl.
Math.(1999), 52, p. 1335-1425



\bibitem{Pa-Sh:07} {\it L. Pastur, M. Shcherbina},  Bulk Universality
and related properties of Hermitian matrix model.-
J.Stat.Phys.(2007), 130, p.205-250

\bibitem{TaoVu:09} T. Tao, V. Vu "Random matrices: universality of local eigenvalue statistics",
arXiv:0906.0510v6 [math.PR]

\bibitem{Er:09}L. Erdos, J. Ramirez, B. Schlein, H.-T. Yau,
"Bulk universality for Wigner hermitian matrices",
arXiv:0905.4176v1 [math-ph]

\bibitem{ErTao:09}L. Erdos, J. Ramirez, B. Schlein, T. Tao, V. Vu, H.-T. Yau,
"Bulk universality for Wigner hermitian matrices with subexponential decay",
arXiv:0906.4400v1 [math.PR]

\bibitem{Mar-Pa:67} {\it V.A. Marchenko, L.A. Pastur,}, Distribution of eigenvalues
for some sets of random matrices. - Math.USSR-Sb.(1967), 1,
p.457-483.

\bibitem{NW:92}
Nagao, T.; Wadati, M. Correlation functions of random matrix ensembles related to classical
orthogonal polynomials. J. Phys. Soc. Japan 61 (1992), 1910Ц1918.

\bibitem{Sil-Ch:95}{\it J.W.Silverstein, S.-I.Choi}, Analysis of the limiting spectral distribution
of large dimensional random matrices. - J. Mult.Anal.(1995), 54, p.295-309

\bibitem{TSh:08} {\it T.Shcherbina}, On Universality of Bulk Local Regime of the Deformed Gaussian
Unitary Ensemble. -preprint

\bibitem{Me:91} {\it M.L.Mehta}, Random Matrices. -Academic Press, New York (1991)

\bibitem{Br-Hi:96} {\it E.Brezin, S. Hikami}, Correlation of nearby levels induced
by a random potential.-Nucl.Phys.(1996),479, p.697-706

\bibitem{Br-Hi:97} {\it E.Brezin, S. Hikami}, Extension of level-spacing
universality.- Phys.Rev. E(1997),56, p.264-269

\bibitem{Br-Hi:98} {\it E.Brezin, S. Hikami}, Level spacing of random matrices in
an external source. -Phys.Rev. E (1998), 58, p. 7176-7185

\bibitem{BaBenPe:05} J.Baik, G.Ben Arours, S.Peche,
Phase transition of the largest eigenvalue for nonnull complex sample covariance matrices".-
Anal. of Prob.(2005), 33,5, p.1643-1697

\bibitem{BenPe:05} G.Ben Arours, S.Peche,
Universality of local eigenvalue statistics for some sample covariance matrices.-
Com. on P. and Apl.Math (2005), LVIII, p.1316-1357

\bibitem{Jo:01}{\it K. Johansson}, Universality of the local spacing distribution
in certain ensembles of Hermitian Wigner Matrices. -Commun. Math.
Phys.(2001), 215, p.683-705

\bibitem{Bl-Ku:04} {\it P.M.Bleher., A.B.J. Kuijlaars},Large n limit of Gaussian
random matrices with external source, part II, - Commun.
Math.Phys. (2004),252, p.43-76

\bibitem{Ap-Co:05} {\it A.I. Aptekarev, P.M. Bleher, A.B.J. Kuijlaars}, Large n
limit of Gaussian random matrices with external source, part II,
-Commun. Math.Phys.(2005),25,p. 367-389

\bibitem{Co-Hi:53}
R. Courant, D. Hilbert, {\em Methods of Mathematical Physics}, Vol. I, Interscience, NY, 1953.

\bibitem{Po-Se:76} G. Polya, G. Szego, {\em Problems and theorems in analysis}.
Die Grundlehren der math.Wissenschaften, Springer-Verlag, Vol. II, 1976 

\end{thebibliography}
\end{document}